\documentclass[12pt]{article}
\usepackage{txfonts}

\usepackage[none]{hyphenat}
\usepackage{graphicx}
\usepackage[rotateright]{rotating}

\title{Genetic code evolution as an initial driving force for molecular evolution}

\author
{Dirson Jian Li* and Shengli Zhang\\
\\
\normalsize{\it Department of Applied Physics, Xi'an Jiaotong
University, Xi'an 710049, China} }

\date{}
\begin{document}

\baselineskip24pt

\setlength{\parskip}{12pt}

\maketitle

\sloppy


{\bf \begin{center} Abstract \end{center}}

There is an intrinsic relationship between the molecular evolution
in primordial period and the properties of genomes and proteomes of
contemporary species. The genomic data may help us understand the
driving force of evolution of life at molecular level. In absence of
evidence, numerous problems in molecular evolution had to fall into
a twilight zone of speculation and controversy in the past. Here we
show that delicate structures of variations of genomic base
compositions and amino acid frequencies resulted from the genetic
code evolution. And the driving force of evolution of life also
originated in the genetic code evolution. The theoretical results on
the variations of amino acid frequencies and genomic base
compositions agree with the experimental observations very well, not
only in the variation trends but also in some fine structures.
Inversely, the genomic data of contemporary species can help
reconstruct the genetic code chronology and amino acid chronology in
primordial period. Our results may shed light on the intrinsic
mechanism of molecular evolution and the genetic code evolution.


\clearpage

\section{Introduction}

The driving force of evolution of life is a core problem in the
theory of evolution. A qualified mechanism on driving force should
explain the evolutionary trends for both molecular evolution and
macroevolution of life. The driving force must be effective
persistently from the primordial period through present days. And it
had to form at the early stage of evolution of life, by which life
evolved from simple to complex consequently. So there must be some
relics in genomic properties of contemporary species resulted from
such a driving force. We found that rich information is stored in
the variation of compositions of proteins and DNAs, which relates to
the evolution in early time. The discovery of genetic code helps us
understand life at the molecular level \cite{Cell
review}\cite{genetic code}\cite{genetic code evol}\cite{genetic code
evol2}\cite{genetic code evol3}. A further study of the evolution of
genetic code may help us reveal the underlying mechanism in the
evolution of life. We found that the genetic code evolution
profoundly determined the evolution of amino acid frequencies and
genomic base compositions, and it can be taken as the initial
driving force in molecular evolution. Inversely, the details of the
genetic code evolution can be inferred by the compositions of
proteins and DNAs of contemporary species.

The organization of the paper is as follows: the experimental
observations and theoretical results on the variation of amino acid
frequencies are explicated in section 2; the experimental
observations and theoretical results on the variation of base
compositions are explicated in section 3; their relationships are
explicated in section 4. In section 5, we will explain the
relationship between genetic code evolution and the variations of
the amino acid frequencies and base compositions. All the
theoretical results in the sections 2-4 are based on a model, which
will be described in details in section 6.

The amino acid frequencies are obtained based on two databases:
($i$) $106$ proteomes ($85$ eubacteria, $12$ archaebacteria, $7$
eukaryotes and $2$ viruses) in Prediction of Entire Proteomes (PEP)
(http://cubic.bioc.columbia.edu/pep) \cite{PEP}, and ($ii$) genomes
of $803$ microbes in the database in NCBI. Two sets of experimental
observations based on PEP and NCBI respectively have been obtained
in the paper. Their results agree with each other. So the properties
on the variation of amino acid frequencies observed in this paper
are universal rules, which is independent of the choice of sample
species. The GC contents are obtained from Genome Properties system
\cite{GC data}. These species are representatives of the three
domains to study the evolutionary trends of amino acid frequencies
and genomic base compositions. Fig. 6a-6d are plotted according to
Fig. 9-1, Fig. 9-6, Fig. 9-8 and Fig. 9-7 in \cite{Forsdyke}
respectively; and Fig. S10a-S10c are plotted according to Fig. 9-4
in \cite{Forsdyke} respectively. The genetic code multiplicity in
Fig. S13 is plotted based on Fig. 1 in Ref. \cite{PRL multiplicity}.
The data of gain-loss of amino acid in Tab. 1 are obtained according
to Ref. \cite{gain_loss}.

\section{Variation of amino acid frequencies}

\subsection{Experimental observations}

\subsubsection{Choosing orders of species properly to observe
variation trends of amino acid frequencies}

The amino acid frequencies in species vary slightly, which was
routinely assumed to be constant \cite{aaf constant}\cite{aaf
constant2}. Unfortunately, this is a misleading assumption and
resulted in ignorance of studying the mechanism of variation of
amino acid frequencies in the past. Actually, it is easy to observe
the variation trends of amino acid frequencies if we choose orders
of species properly. In this section, several orders of species will
be introduced based on the amino acid chronology or the like. Thus,
we can obtain the variation trends of amino acid frequencies, which
can help us reveal the mechanism of the variation of amino acid
frequencies in the evolution.

The chronology of amino acids to recruit into the genetic code from
the earliest to the latest can be estimated as: G, A, D, V, P, S, E,
L, T, R, Q, I, N, H, K, C, F, Y, M, W \cite{AA Chronology}. Let
$a(i),\ i=1...20$ denote the $20$ amino acids in this chronological
order. According to this amino acid chronology, some amino acids
such as G and V recruited into the genetic code earlier than other
amino acids such as H, Q and W. Let
$$f(a(i),\xi)=\frac{N(a(i),\xi)}{\sum_{j=1}^{20}N(a(j),\xi)},\
\xi=1,...,n_s$$ denote the frequency of amino acid $a(i)$ for the
species in PEP ($n_s=106$) or in NCBI ($n_c=803$), where
$N(a(j),\xi)$ denotes the total number of amino acid $a(j)$ in all
the protein sequences of species $\xi$. If we sort the species
properly, we can observe variation trends of amino acid frequencies
when $f(a(i),\xi)$, roughly speaking, increases or decreases with
$\xi$.

We can introduce Late-early Ratio Orders to sort the species
properly to observe variation trends of amino acid frequencies. In
this class of orders, we can arrange the species by
$R_{(a(i)...a(j)) / (a(k)...a(l))}(\xi)$, namely the ratio of the
average amino acid frequency of $f(a(i),\xi),...,f(a(j),\xi)$ to the
average amino acid frequency of $f(a(k),\xi),...,f(a(l),\xi)$, where
$a(i),...,a(j)$ are some later recruited amino acids and
$a(k),...,a(l)$ some earlier recruited amino acids. The Late-early
Ratio Order is generally chronological in the evolution according to
its definition. When the earlier recruited amino acids or later ones
are given concretely, we can define
$$R_{10/10}(\xi)=\frac{\sum_{i=11}^{20}f(a(i),\xi)}{\sum_{i=1}^{10}
f(a(i),\xi)}$$ and obtain $R_{10/10}$ order. Similarly, we obtain
$R_{HQW/GV}$ order and $R_{1/G}$ order, where
$$R_{HQW/GV}(\xi)=\frac{\frac{1}{3}\sum_{a=H, Q, W}f(a,\xi)}{\frac{1}{2}\sum_{a=G,
V}f(a,\xi)},$$ and $$R_{1/G}(\xi)=\frac{1}{f(G,\xi)}.$$ $R_{10/10}$
order, $R_{HQW/GV}$ order and $R_{1/G}$ order are some cases of
Late-early Ratio Orders.

If we choose orders of species improperly, the variation trends of
amino acid frequencies can not be observed. An improper choice is
the Random Ratio Order. In this order, we arrange the species by
Random Ratio Order $R_{(a(i)...a(j)) / (a(k)...a(l))}(\xi)$, where
the amino acids in numerator and denominator are chosen randomly
among the $20$ amino acids. By this way, we can not observe
variation trends of amino acid frequencies when $f(a(i),\xi)$ vary
with $\xi$ randomly. For instance, the species can be arranged by
$R_{AGHCN/LVQW}$ order, where
$$R_{AGHCN/LVQW}(\xi)=\frac{\frac{1}{5}\sum_{a=A, G, H, C, N}f(a,\xi)}
{\frac{1}{4}\sum_{a=L, V, Q, W}f(a,\xi)}.$$

At last, we can introduce $L_{av}$ order to observe the variation
trends of amino acid frequencies roughly. In this order, the species
can be arranged from short to long by the average protein length
$L_{av}(\xi)$ of all the proteins in species. The $L_{av}$ order is
independent of the choice of amino acids according to its
definition.

\subsubsection{General variation trends of amino acid frequencies}

When sorting the species properly, we can obtain a set of variation
trends for the $20$ amino acids. The variation trends are generally
common for different proper orders of species, and the results of
variation trends are also common either based on the data of species
in PEP or based on the data of species in NCBI. So the general
variation trends of amino acid frequencies are intrinsic properties
of species.

When sorting species by $R_{10/10}$ order, the amino acid
frequencies based on $106$ species in PEP is in Fig. 1a, and the
result based on $803$ species in NCBI is in Fig. S4. Both of the
results are the same in variation trend for each of the $20$ amino
acids. Roughly speaking, the frequencies of G, A, D, V, P, L, T, R,
H, W tend to decrease, the frequencies of S, E, I, N, K, F, Y tend
to increase, and the frequencies of Q, C, M tend to keep constant.
The magnitudes of variations are different: frequencies of G, A, V,
P, R decrease more rapidly than that of D, L, T, H, W, while
frequencies of I, N, K, F, Y increase more rapidly than that of S,
E. We found that the evolutionary trends of amino acids are related
to the amino acid chronology \cite{AA Chronology}: most of the amino
acids whose frequencies tend to decrease (or increase) are among the
earlier (or later) recruited amino acids according to the amino acid
chronology. The variation trends of amino acid frequencies are
intrinsic properties of molecular evolution, which are irrelative to
the choice of orders in sorting species. We can observe generally
the same variation trends by $R_{10/10}$ order, $R_{HQW/GV}$ order,
$R_{1/G}$ order and $L_{av}$ order (Fig. 1a and Fig. S1). The
variation trends are irrelative to the choice of amino acids,
because we can also observe the same variation trends by $L_{av}$
order (Fig. S1c and Fig. S2). If sorting the species improperly, we
can only observe completely random variation of amino acid
frequencies (Fig. S3).

\subsubsection{Fine structures of the variation of amino acid frequencies}

Furthermore, we can observe fine structures and superfine structures
of the variation of amino acid frequencies because $f(a(i),\xi)$ do
not vary with $\xi$ linearly in general. We choose $R_{10/10}$ order
in studying the variation of amino acid frequencies, which is an
intrinsic order related to the molecular evolution. Roughly
speaking, the species with lower (or higher) series number by
$R_{10/10}$ order may appear earlier (or later) in the evolution
according to the definition of $R_{10/10}$. The details of variation
of amino acid frequencies can be observed more obviously when
studying the data of $803$ species in NCBI than studying the data of
$106$ species in PEP. In the research, we obtained smoothed lines of
the discrete dots ($\xi$, $f(a(i),\xi)$) for each amino acid
according to Savitzky-Golay method \cite{Savitzky-Golay}. The fine
structures and superfine structures of the variation of amino acid
frequencies can be observed explicitly according to the fluctuations
in the smoothed lines.

We prescribe the fine structures as the profiles of the variations
of amino acid frequencies. The smoothed lines corresponding to fine
structures can be obtained by certain greater spans according to
Savitzky-Golay method (black lines in Fig. 2). The profiles of dots
of the amino acid frequencies with respect to series numbers of
species are S-shaped or inverse S-shaped in general. For examples,
the profile of the $803$ dots ($\xi$, $f(V,\xi)$) is S-shaped and
the profile of the $803$ dots ($\xi$, $f(F,\xi)$) is inverse
S-shaped (see subplot V and subplot F in Fig. 2). Generally
speaking, we can observe more steep variation trends at both ends.
In the case for amino acids G, A, D, V, P, T, R whose variation
trends are decreasing, the smoothed lines go down more quickly at
both left and right, so we can observe S-shaped profiles
approximately. In the case for amino acids S, I, N, K, F, Y whose
variation trends are increasing, the smoothed lines go up more
quickly at both ends, so we can observe inverse S-shaped profiles
approximately. In the case for other amino acids Q, C, M, L, H, W,
E, we can observe deformed S-shaped or waved profiles. Especially,
we found that the trends are always well S-shaped or inverse
S-shaped for most of the amino acids whose average amino acid
frequencies are high, while the trends are deformed S-shaped for the
amino acids whose average amino acid frequencies are low.

\subsubsection{Superfine structures of the variation of amino acid frequencies}

We prescribe the superfine structures as the fluctuations in the
profiles of the variations of amino acid frequencies. The smoothed
lines corresponding to superfine structures can be obtained by
certain smaller spans according to Savitzky-Golay method (red lines
in Fig. 2). We also sorted the species by $R_{10/10}$ order. We can
show that the superfine structures do exist in the background of
stochastic fluctuations. We can observe some obvious fluctuations in
the smoothed lines. For examples, we can observe an obvious waved
characteristic of fluctuations for amino acid W (for series number
of species around $300$ in the subplot W in Fig. 2); we can observe
a peak for amino acid S (for series number of species around $500$
in the subplot S in Fig. 2); we can observe two peaks for amino acid
M (for series number of species around $400$ and $600$ in the
subplot M in Fig. 2). Some obvious superfine structures can be
observed in either the result based on NCBI (Fig. 2) or the result
based on PEP (Fig. S2). We will explain the intrinsic property of
the superfine structures by a model in section 2.2.3.

\subsubsection{Variation trends of amino acid frequencies for three domains}

It is significant to compare the variation trends of amino acid
frequencies for three domains (eubacteria, archaebacteria, and
eukaryotes). The evolution trends of each of the $20$ amino acid
frequencies are the same without exception for three domains (Fig.
4). But there are differences of the ranges of amino acid
frequencies for three domains. The ranges of the variations of amino
acid frequencies for eubacteria, roughly speaking, coincide with the
ranges of the variations of amino acid frequencies for
archaebacteria. There are, however, obvious differences between the
ranges for archaebacteria and eukaryotes (Fig. 4). Namely, the
initial amino acid frequencies for start series number of eukaryotes
by $R_{10/10}$ order are less than (or greater than) initial amino
acid frequencies for start series number of archaebacteria by
$R_{10/10}$ order for amino acids whose trends decrease (or
increase).

\subsection{Theoretical results}

\subsubsection{General variation trends of amino acid frequencies}

The variation trends of amino acid frequencies simulated by the
model (Fig. 1b and red dots in Fig. S4) agree with the variation
trends in experimental observations (Fig. 1a and celeste dots in
Fig. S4) very well. The mechanism on the variation of amino acid
frequencies is faithfully based on the genetic code multiplicity.
The variation trends and magnitudes of amino acid frequencies are
crucially related to the placements of amino acids in the genetic
code multiplicity (Fig. S13). For examples, the amino acids F and Y
occupy similar positions in the genetic code multiplicity, so the
corresponding evolutionary trends and magnitudes accord with each
other. The similar results are also valid for the amino acids G and
A, for the amino acids P and H, for the amino acids R, D and E and
for the amino acids L and S respectively (Fig. 1, S4 and S13).
Hence, our result reveals that the underlying mechanism of the
variation of amino acid frequencies in the contemporary species is
determined by the evolution of genetic code.

The magnitudes of the variations of amino acid frequencies in the
theoretical results are about half of the corresponding experimental
observations. We did not add any more parameters in the model to
boost the magnitudes of variation of amino acid frequencies in the
model so as to obtain a better result concerning to the magnitudes,
because we insisted on that the model is faithfully based on the
genetic code multiplicity.

\subsubsection{Fine structures of the variation of amino acid frequencies}

Up to a certain precision, the variation trends in theoretical
results are linear for all the amino acids (red lines in Fig. S4).
But the fine structures of the variation of amino acid frequencies
obviously deviate from linear relationships in experimental
observations (blank lines in Fig. S4 or blank lines in Fig. 2 for
zoom in). According to the model, we can easily explain the S-shaped
or inverse S-shaped profiles in fine structures of the variation of
amino acid frequencies. The number of species in NCBI varies with
average protein length as in Fig. S8c. This distribution indicates
that the sample density of species in NCBI is not a constant when
considering the relationship between amino acid frequencies and
average protein length in Fig. 8. So series number does not vary
linearly with the time in the evolution. Namely we only obtained
less samples of species in the sections with small and big series
numbers by $R_{10/10}$ order than in the sections with medium series
numbers by $R_{10/10}$ order. As a result, we can observe more steep
variation trends in the sections with small and big series numbers
in the fine structures.

In the simulation in Fig. S4, we choose a linear relationship
between parameter $t$ and series number of species $N_s$, i.e.,
$t=t_1(N_s)$, without consideration of the variation of sample
density. We introduce an simple function $t=t_3(N_s)$ so as to
obtain a similar relationship between number of species and
parameter $t$ (Fig. S8a and S8b). Thus we obtained the fine
structures of variations of amino acid frequencies (blank lines in
Fig. 3, and Fig. S7 for zoom out). If we choose an improper function
$t=t_2(N_s)$, which corresponds to unreasonable sample density, the
simulated fine structures can not agree with the experimental
observations in the sections with small series numbers (left ends of
lines in Fig. S6). Therefore, we show that the reason to be either
S-shaped profiles or inverse S-shaped profiles is trivially due to
sample density of species in the databases.

The results in simulation (blank lines in subplots Q, C, M, L, H, W,
E in Fig. 3) can not agree well with the deformed S-shaped
experimental observations (blank lines in corresponding subplots in
Fig. 2). If we can understand the evolution of genetic code in more
details, the model may be improved and we may obtain a better result
in the future.

\subsubsection{Superfine structures of the variation of amino acid frequencies}

We can observe the superfine structures of the variation of amino
acid frequencies in theoretical results (red lines in Fig. 3). In
the simulation, the amino acid frequencies are calculated based on
sufficient numerous simulated protein sequences so as to avoid the
random errors. All the fluctuations in Fig. 3 can recur in details
if we run the program again. Therefore, the fluctuations of these
smoothed lines are intrinsic properties of the variation of amino
acid frequencies, which are determined by the genetic code
multiplicity according to the mechanism of the model.

The simulation results indicate that the superfine structures in the
experimental observations are also intrinsic properties (not just
noises in observations) and should also be determined by the genetic
code multiplicity. In this case, we can not compare the fluctuations
between theoretical results and experimental observations in
details. However, this heuristic conclusion may help us understand
the evolution of genetic code in details by studying the
fluctuations of superfine structures of the variation of amino acid
frequencies.

It is possible for us to compare characteristics of superfine
structures between theoretical results and experimental observations
by the model. In the previous result in Fig. 2, we have introduced
an artificial function $t=t_3(N_s)$. If we still let $t$ vary
linearly with $N_s$, i.e., $t=t_1(N_s)$, the results will only based
on the genetic code multiplicity (red lines in Fig. S4). After
subtracting the linear parts (in the sense of least squares), we can
obtain the intrinsic fluctuations (not noises) of variation of amino
acid frequencies (Fig. S5). We can also observe an obvious waved
characteristic of fluctuations for amino acid W (subplot W in Fig.
S5) and an double-peak characteristic for amino acid M (subplot M in
Fig. S5), which might agree with the experimental observations.

\subsubsection{Variation trends of amino acid frequencies for three domains}

According to the phylogeny of three domains, Eukarya and Archaea
separated later in the evolution, while Bacteria and the ancestor of
both Eukarya and Archaea separated earlier in the evolution (Fig.
S9) \cite{3domains}. Hence, we can explain the differences of
variation of amino acid frequencies for three domains in
experimental observations in terms of the phylogeny of three
domains. The unicellular organisms appeared earlier and eukaryotes
appeared later in the history. We input some initial values of
$iAAF_{\mbox{\tiny NCBI}}$ (Tab. 2) firstly in simulations of the
evolution of amino acid frequencies for Bacteria and Archaea. Then,
we input the medium amino acid frequencies $iAAF_{\mbox{\tiny
medium}}$ of the previous step as initial amino acid frequencies to
simulation the evolution of amino acid frequencies for Eukarya (Fig.
S9). The simulation results (Fig. 5) agree with the experimental
observations (Fig. 4) in general.

\section{Variation of genetic code compositions}

\subsection{Experimental observations}

It is well known that base compositions in genomes vary greatly,
which are often referred as GC pressure or GA pressure in molecular
evolution. There are delicate structures in the variation of genomic
base compositions of contemporary species, which are discussed in
details in Ref. \cite{Forsdyke}.

The precise correlations between genomic GC content (or genomic GA
content) and the GC content (or GA content) at the first, second, or
third codon positions can be observed obviously (Fig. 6a, 6b)
\cite{GC GC123} \cite{Forsdyke}. There are also correlations between
codon position GC contents and codon position GA contents (Fig. 6c),
or between genomic GC content and genomic GA content (Fig. 6d)
\cite{Forsdyke}. Moreover, there are correlations between GC content
of genes and codon position GC contents of genes in each genome,
hence we can obtain three slopes of the corresponding correlations
in first, second and third codon positions for a species
\cite{Forsdyke}. The slopes corresponding to the three codon
positions vary with genomic GC content respectively for contemporary
species (Fig. S10a-S10c) \cite{Forsdyke}.

\subsection{Theoretical Results}

The mechanism of the evolution of genomic base compositions can also
be explained by the same model based on genetic code multiplicity
and codon chronology. The simulations of correlations of genomic
base compositions and codon position base compositions (Fig. 6e-6h)
agree with the experimental observations respectively (Fig. 6a-6d).
It is noteworthy that there are many detailed agreements between
simulations and experimental observations, which strongly confirm
the validity of our simulations.

In Fig. 6e, we can observe a step in the middle of the line
corresponding to the first codon position and a junction between
lines corresponding to the first and second codon positions; these
characters of step and junction can also be observed in the plots
based on biological data \cite{GC GC123}\cite{Forsdyke}\cite{GC
GC123 physica a}. The slope of the line corresponding to the third
codon position is the deepest, because G and C occupy all the third
positions of the earliest codons for $20$ amino acids, and A and U
occupy all the third position of the latest codons for $20$ amino
acids, but their compositions are about invariant for the first and
second positions (Tab. 5-6). The lower limit and upper limit of the
GC content for contemporary species also result from the base
compositions in codon positions in a chronological list of codons.
In Fig. 6f, the simulated slope corresponding to the third codon
position is the greatest, which agrees with the experimental
observation \cite{Forsdyke}. In Fig. 6g, the slopes and variation
range in simulation agree with the experimental observation
\cite{Forsdyke}. And in Fig. 6h, the deviation amplitude from the
central declining line of the correlation between genomic GC content
and genomic GA content is great, which agrees with the big standard
error in Fig. 9-7 in Ref. \cite{Forsdyke}. At last, the simulations
of the correlation between genomic GC content and the three slopes
of correlations of GC content in genes (Fig. S10d-S10f) agree with
the experimental observations \cite{Forsdyke} in principle. Thus, we
show that the delicate structure in the correlations of genomic base
compositions mainly comes from genetic code multiplicity and
chronology (Fig. 6i-6l) \cite{PRL multiplicity}\cite{genetic code
chronology}.

\section{Relationships among amino acid frequencies, genomic base
compositions and average protein lengths}

\subsection{Experimental observations}

\subsubsection{Relationships between amino acid frequencies and
genomic base compositions}

There is explicit relationship between amino acid frequencies and
genomic base compositions. We observe that the genomic GC content
decreases linearly with the ratio $R_{10/10}$ (Fig. 7a) \cite{aaf
GC}\cite{aaf GC2}. Similar results are also valid for other
Late-early Ratio Orders. So the genomic GC content and the amino
acid frequencies are not independent variables when we discuss the
evolutionary pressure in molecular evolution. It is often taken
genomic GC content as the initial evolutionary pressure. If we can
explain the relationships among amino acid frequencies, genomic base
compositions and average protein lengths all together, it becomes
not reasonable to take genomic GC content as the initial
evolutionary pressure.

\subsubsection{Relationships between amino acid frequencies and average protein lengths}

There is also correlation between the average protein length
$\bar{l}$ and the ratio $R_{HQW/GV}$ (Fig. 8). The distribution of
all species forms a bowed line in the $\bar{l} - R_{HQW/GV}$ plane,
and the closely related species cluster together in the $\bar{l} -
R_{HQW/GV}$ plane (Fig. 8). Roughly speaking, the average protein
length becomes longer and longer in the evolution and there will be
more and more opportunities for later recruited amino acid to appear
in the protein sequences. Similar results are also valid for other
Late-early Ratio Orders. We choose $R_{HQW/GV}$ in order that the
distribution of archaebacteria can be separated from the
distribution of bacteria. We also noticed that the species with
larger genome size locate in the midstream of the evolutionary flow.
So this bowed distribution can be interpreted as an evolutionary
flow. And more advanced prokaryotes with larger genome sizes always
locate in the midstream of this evolutionary flow.

\subsection{Theoretical Results}

\subsubsection{Relationships between amino acid frequencies
and genomic base compositions}

According to the simulation, we found that the genetic code
multiplicity can influence both amino acid frequencies and genomic
GC content. So the evolutionary pressure in the overall molecular
evolution originated in the genetic code evolution. The evolutionary
pressure influences the amino acid frequencies, genomic base
composition and the average protein length in proteome all together.

When the parameter $t$ increases, there will be more and more late
amino acids to join the protein sequences, so the ratio $R_{10/10}$
will also increase, but the genomic GC content will decrease
according to the codon chronology. Thus we can explain that the
ratio $R_{10/10}$ increases when the genomic GC content decrease
(Fig. 7b). The variation range of $R_{10/10}$ in simulation (Fig.
7b) is less than the variation range of $R_{10/10}$ in experimental
observation (Fig. 7a), which is due to that the magnitudes of
variations of amino acid frequencies are insufficient in simulations
by the model.

The variation of amino acid frequencies also influenced the genomic
base compositions. For example, the total numbers of bases G and C
are equal to $8$ in all stages for the second codon position and
almost constant ($9$ to $10$) for the first codon position (Tab. 6
and Fig 6i). In the simulation result, however, the GC content at
the first and second codon position obviously decrease from right to
left (corresponding to parameter $t$ increases from $t_{min}$ to
$t_{max}$) (Fig. 6e). In the simulation, the probabilities for the
late recruited amino acids (with less bases G and C in their codons,
see Tab. 5) to join the proteins will increase when the parameter
$t$ increases, therefore the GC content at the first and second
codon positions decrease from right to left in Fig. 6e. Thus, we can
explain the slopes of the correlations between genomic GC content
and the GC content at codon positions in Fig. 6a.

\subsubsection{Relationships between amino acid frequencies and average protein lengths}

The relationship between amino acid frequencies and average protein
length can be explained by the model. The distribution of species in
the $\bar{l} - R_{HQW/GV}$ plane can be simulated by our model (Fig.
8, Embedded). The bending direction of the simulated evolutionary
flow agrees with the evolutionary flow in experimental observation.
According to the simulation by the model, the bending direction
sensitively relates to the genetic code multiplicity. If we vary the
substitution rules a little, the bending curve may disappear in the
result by simulation and we can only obtain a linear relationship
between $R_{HQW/GV}$ and $\bar{l}$. So such a detail agreement
between theoretical result and experimental observation verifies the
validity of the referred genetic code multiplicity in the model
(Fig. S13 and Ref. \cite{PRL multiplicity}).

\section{Genetic code evolution as an initial driving force for molecular evolution}

\subsection{Timing the mechanism of the variation of amino acid frequencies}

The estimation of the time when the variation of amino acid
frequencies appeared firstly will help us seek the mechanism that
determines the variation of amino acid frequencies of contemporary
species. We can explain that the variation of amino acid frequencies
formed before the stage when three domains began to branch.

Firstly, we explain the incorrelation between the variation of amino
acid frequencies in our observations and the gain-loss of amino
acids in modern time. The trend of amino acid gain and loss in
protein evolution has been reported in Ref. \cite{gain_loss}. There
is a debate on the mechanism of the variation of amino acid
frequencies \cite{gain_loss debate}. There is a set of data of gain
and loss for $20$ amino acids in Ref. \cite{gain_loss} according to
the protein evolution in modern time (Tab. 1). We can also obtain a
set of data of variation trends for $20$ amino acids according to
Fig. 1a by least squares (Tab. 1). The correlation efficient between
gain-loss and variation trends is $0.393$, which means that the
variation trends are incorrelated with the gain and loss of amino
acids in modern time. If it is true, we can infer that the mechanism
of the variation of amino acid frequencies should appeared in early
period.

The variation of amino acid frequencies for three domains may help
us time the mechanism of the variation of amino acid frequencies.
The evolutionary trends of amino acid frequencies are the same for
three domains (Fig. 4). And we have explained the differences of
amino acid frequencies for three domains according to the phylogeny
of three domains in section 2.2.4. These results indicate that the
mechanism of the variation of amino acid frequencies should
originate in the period before the separation between Bacteria and
the ancestor of Archaea and Eukarya.

\subsection{Initial driving force for molecular evolution}

The variations of amino acid frequencies and genomic base
compositions can be explained in a unified theoretical framework
based on genetic code evolution. All the theoretical results in
section 2-4 are based on one model, whose core is the genetic code
multiplicity and chronology. The simulations by the model agree with
the experimental observations not only in variation trends but also
in many detailed characteristics. So there is close relationship
between the variation of amino acid frequencies and genomic base
compositions of contemporary species and the genetic code evolution
in the primordial period. We believe that ($i$) the pattern of the
variation of the compositions of protein and DNA formed and fixed in
the period when the genetic code evolved; ($ii$) the magnitudes of
the evolutionary trends have been amplifying ever since the genetic
code had established. Thus, we conjecture that the genetic code
evolution is an initial driving force for molecular evolution.

\subsection{Cracking the genetic code evolution}

If our conjecture is reasonable, we found a new method to crack the
genetic code evolution in primordial time by studying the variation
of amino acid frequencies and genomic base compositions of
contemporary species. Especially, there are some detailed
disagreements between theoretical results and experimental
observations. For instance, we can not compare the superfine
structures in details between theoretical results and experimental
observations. The improvement in understanding the variation of
amino acid frequencies and genomic base compositions may lead us to
crack the genetic code evolution.

\section{The model}

\subsection{Outline of the model}

We proposed a model based on the evolution of genetic code to
explain the variations of amino acid frequencies and base
compositions in species. The model mainly consists of three parts:
($i$) generating void protein sequences by formal language; ($ii$)
generating protein sequences based on genetic code multiplicity; and
($iii$) generating DNA sequences based on codon chronology (Fig.
S11).

The genetic code multiplicity (Fig. 13) is the core in simulation of
variation of amino acid frequencies. The genetic code chronology
(Tab. 5) is the core in simulation of variation of genomic base
compositions. This is an elaborate model and there is only one
adjustable parameter $t$, which indicates the time of the evolution
and plays central roles in simulations of the variations of amino
acid frequencies and base compositions.

\subsection{Simulation of the variation of amino acid frequencies}

\subsubsection{Generation of void protein sequences by formal language}

There are two steps in generation of protein sequences in the model:
(1) generating void protein sequences according to tree adjoining
grammars in Fig. S12 \cite{TAG}; (2) the leaf $\pi$ in the tree
adjoining grammar will be substituted by amino acids based on the
genetic code multiplicity in Ref. \cite{PRL multiplicity} (Fig. S13
and Tab. 3), where the amino acid chronology has been considered
according to Ref. \cite{AA Chronology} and the probabilities for
substitutions are determined by $p_{a}$ in Tab. 2.

There are no rigorous restrictions in choosing grammar rules of the
formal language, because they do not essentially determine the
variation trends of amino acid frequencies in our final results. In
the model, we choose tree adjoining grammar to generate void protein
sequences. There are one initial tree and two associate trees in the
grammar rules, where $S$ and $T$ are inner nodes and $\pi$ is leaf
(Fig. S12). We start from the initial tree when generating a void
protein sequence. Then the inner nodes $S$ or $T$ can be substituted
by the corresponding associate trees (see the example in Fig. S14).
The void protein sequence consists of the leaves around the final
tree.

We set a parameter $t$ in the model as the probability of the
substitution of inner nodes. The length of a sequence may increase
if an inner node has been substituted by the corresponding auxiliary
tree (with probability $t$) or keep constant if the node has not
been substituted (with probability $1-t$). The void protein
sequences consist of void character $\pi$, which will be substituted
by amino acids in the next step. When the parameter $t$ is fixed, we
can generate a group of void sequences with certain length
distribution. In the model, the void protein sequences increase only
several units in length in each principal cycle of the program. The
number of cycles in the program is constant in the model. We can
obtain longer void protein sequences after running more principal
cycles. The average increment of protein length in one principal
cycle varies with respect to parameter $t$. When $t$ increases,
there will be more probability to generate longer void protein
sequences in each principal cycle.

\subsubsection{Fill in the protein sequences based on genetic code multiplicity}

In the second step, the void character $\pi$ in the void protein
sequences will be substituted by $20$ amino acids by calling a
subprogram. The rules of substitutions in the subprogram are based
rigorously on the genetic code multiplicity (Fig. S13, Tab. 3).
Namely, $\pi$ may be substituted by either of $\pi_1$, $\pi_2$,
$\pi_3$, $\pi_4$, $\pi_5$ or amino acid $C$ at first level of
substitution; $\pi_1$ may be substituted by either of $R$, $E$, $D$
or $\pi_6$ at second level of substitution, and so on; $\pi_6$ may
be substituted by either of $H$ or $P$ at third level of
substitution, and so on (Fig. S13, Tab. 3). The depth of
substitutions of the genetic code multiplicity tree is four levels
in maximum, namely from $\pi$ to $N$, $Q$, $W$ or $M$ (Fig. S13,
Tab. 3). The process of substitution of $\pi$ will not finish until
$\pi$ is substituted by one of the $20$ amino acids eventually.
Thus, the protein sequences consisting of amino acids have been
generated.

The probabilities for all the possible substitutions of each node in
the genetic code multiplicity tree (Fig. S13, Tab. 3) are constant
parameters in the model according to the average amino acid
frequencies of contemporary species. The initial amino acid
frequencies (iAAF) at the beginning of each cycle in the program are
input by the average amino acid frequencies of species in PEP
($n_s=106$) or NCBI ($n_s=803$) (Tab. 2):
$$f_{a(i)}=\frac{\sum_{\xi=1}^{n_s}N(a(i),\xi)}{\sum_{j=1}^{20}
\sum_{\xi=1}^{n_s}N(a(j),\xi)}.$$ We choose the average amino acid
frequencies as the initial amino acid frequencies in order that the
average amino acid frequencies in the simulation results agree with
the average amino acid frequencies in observation. But these
constant parameters do not influence the variation trends in the
simulations. The corresponding probabilities for $20$ amino acids
can be calculated as followings:
$$p_{a(i)}=\frac{f_{a(i)}}{\sum_{i=1}^{20}f_{a(i)}+f_{term}},$$ where $f_{term}=f_I$
(namely $\pi_{10}$ will be substituted by $\pi_{11}$ and $I$ with
the equal probability in Fig. S13). According to Tab. 3, all the
probabilities of substitution between two connected nodes in Fig.
S13 can be calculated by the initial amino acid frequencies. In the
substitution of $\pi_6$ by $H$, for instance, the probability is
$p_H/(p_H+p_P)$; while the probability of substitution of $\pi$ by
$\pi_2$ is $p_2=p_F+p_Y$ (Tab. 2).

\subsubsection{Calculation of amino acid frequencies and their variation trends}

When the parameter $t$ is fixed, we can generate sufficient numerous
protein sequences so that the amino acid frequency for each of the
$20$ amino acids goes to a certain constant value. Then, the amino
acid frequencies vary with parameter $t$.

We can study the evolution of amino acid frequencies by adjusting
the parameter $t$ from $t_{min}$ to $t_{max}$. We observed that the
amino acid frequency for each amino acid may increase or decrease
when $t$ increases. For the initial parameter $t_{min}$, we input
$p_{a(i)}$ as initial values of amino acid frequencies. We chose
$n_m$ sample species $N_s=1, 2, ..., n_m$ and obtained $n_m$ sets of
amino acid frequencies by the model. Hence, we can obtain the
variation trends of amino acid frequencies in the simulation. We
found that the variation trends of amino acid frequencies are
irrelative to the initial values of the amino acid frequencies,
which are determined by the genetic code multiplicity.

\subsubsection{The mechanism of the variation of amino acid
frequencies in the simulation}

According to the simulation by the model, the genetic code
multiplicity plays central role in the evolution of amino acid
frequencies. It should take more steps of substitutions to replace
$\pi$ in the void protein sequences by a late recruited amino acid.
The placements of amino acids in the genetic code multiplicity tree
essentially influence the variation of amino acid frequencies in the
simulation. When we generate shorter protein sequences, the late
recruited amino acids have less opportunities to join the protein
sequences. When the parameter $t$ increases, the increment of
protein length in one principal cycle also becomes greater than
before. Hence, there will be more opportunities for the late
recruited amino acids to join the protein sequences. So the amino
acid frequencies will vary with the parameter $t$. The only
continuous variable $t$ in the model can be interpreted as the time
in the evolution. The other constant parameters $p_{a(i)}$ and the
grammar rules do not essentially influence the variation trends. And
there are no other parameters in the model to deliberately influence
certain amino acid frequencies or genomic base compositions.

\subsection{Simulation of the variation of genomic base
compositions}

\subsubsection{Codon chronology}

The codon chronology can be reconstructed based on the amino acid
chronology and the primacy of thermostability and complementarity
(Tab. 4) \cite{AA Chronology}\cite{genetic code
chronology}\cite{genetic code chronology2}\cite{complementarity}.
The result of codon chronology in Tab. 4 is almost the same as the
chronology in Ref. \cite{genetic code chronology2}, but amino acid
chronology in the first line of Tab. 4 in our calculation is
replaced by a new amino acid chronology obtained by the same authors
in Ref. \cite{AA Chronology} that was published later than Ref.
\cite{genetic code chronology2}. In our results, the amino acids are
sorted chronologically in the first line in Tab. 4. The $32$ pairs
of complementary codons are sorted chronologically from top to
bottom in Tab. 4, which correspond to the above amino acid
chronology and form a lower triangle in Tab. 4.

The codon chronology in Tab. 4 can explain the relationship between
GC content and codon position GC content very well. But the
declining relationship between genomic GC content and genomic GA
content in experimental observations (Fig. 6d) must not be achieved
by the codon chronology in Tab. 4 in principle because of the
rigorous restriction of codon complementarity. We have to modify the
chronology slightly in sacrifice of the codon complementarity so
that the relationship between genomic GC content and genomic GA
content in experimental observation (Fig. 6d) can be simulated by
the model. We only rearranged the codon chronology for amino acids S
and R and obtained a modified codon chronology (Tab. 5). Namely, we
moved the positions of codons AGC and AGU earlier in Tab. 5 than in
Tab. 4 for amino acid S; and we moved the positions of codons AGG
and AGA earlier in Tab. 5 than in Tab. 4 for amino acid R.

According to the codon chronology in Tab. 5, there are $32$ stages
in the evolution of genetic codes. For each stage, there is a
definitive correspondence between a codon and a certain amino acid.
For each stage, therefore, we can count the total numbers of bases
G, C, A or T for all the amino acids in the first, second or third
codon positions respectively according to Tab. 5 (Tab. 6). And we
also obtained the total number of G and C and the total number of G
and A in the first, second and third codon positions respectively as
well as the total numbers of G and C or G and A in all the three
codon positions (Tab. 6). The results in Tab. 6 partially indicate
the variation trends of genomic base compositions. And we can plot
the relationships in Fig. 6i-6l according to Tab. 6.

\subsubsection{Generation of DNA sequences according to codon chronology}

There is a one to one relationship between amino acids and a codon
at each of the $32$ stages according to the codon chronology in Tab.
5. We separated the section [$t_{min}$, $t_{max}$] equally into $32$
subsections so that each value of parameter $t$ is in one of these
subsections. Thus, the amino acids in the protein sequences
generated in the second part of the model can be replaced by
corresponding codons at certain stages. So there is also a one to
one correspondence between protein sequences and DNA sequences at
each stage. Thus a group of DNA sequences can be obtained in the
model as soon as the protein sequences have been generated, which is
based on both the genetic code multiplicity and the codon
chronology.

\subsubsection{Calculation of genomic base compositions and their variation trends}

For a fixed value of parameter $t$, which corresponds to a certain
stage, we can obtain numerous DNA sequences. Then we can calculate
the compositions of bases G, C, A, T. At last, we can obtain GC
content or GA content at certain codon positions and the total GC
content or GA content in all the DNA sequences. When the parameter
$t$ varies, we can study the evolution of genomic base compositions
or base compositions at certain codon positions.

\subsubsection{The mechanism of the variation of genomic base
compositions in the simulation}

According to the simulation in the model, the variation of base
compositions in contemporary species can be explained by the codon
chronology, genetic code multiplicity and amino acid chronology
together. For instance, we can explain the relationship between
genomic GC content and GC content at the first, second and third
codon positions. The plot of the relationship between total GC
numbers and GC number at certain codon positions in Fig. 6i is
approximate the same as the corresponding experimental observation
in Fig. 6a. The upper limit (about $25\%$) and lower limit (about
$75\%$) of genomic GC content are due to the abundance of G and C in
the earliest codons and lack of G and C in the latest codons in Tab.
5. The characteristic of ``step'' at the middle for the third codon
position in experimental observation is due to the leap of numbers
of G and C at third codon position in Tab. 6. There are always $8$ G
and C at the second codon position in Tab. 6, so there is no step
for the second codon position in experimental observation. The
simulation of GC content for each codon position increases with
genomic GC content (Fig. 5e), which agrees with the experimental
observation (Fig. 5a). When $t$ increases, there will be less
opportunities for bases G and C to join the DNA sequences. The other
relationships in Fig. 6 can also be explained similarly.

\subsection{Simulation of relationships among amino acid frequencies, genomic base
compositions and average protein lengths}

\subsubsection{Relationships between amino acid frequencies and
genomic base compositions}

When the parameter $t$ is fixed, both of a group of proteins and a
group of corresponding DNA sequences can be generated together by
the model. Subsequently, both of the amino acid frequencies and
genomic GC content can be calculated. Thus we can explain that the
ratio $R_{10/10}$ increases when the genomic GC content decrease
(Fig. 7a).

\subsubsection{Relationships between amino acid frequencies and
average protein lengths}

When the parameter $t$ is fixed, a group of proteins can be
generated by the model. So both the amino acid frequencies and
average protein length can be calculated together. When $t$ varies,
we can obtain the relationship between $R_{HQW/GV}$ and average
protein length $\bar{l}$ (Fig. 8, Embedded), which agrees with the
experimental observation very well (Fig. 8). We obtained a bended
curve for the relationship between $R_{HQW/GV}$ and $\bar{l}$, whose
bending direction is upward (Fig. 8, Embedded). The bending
direction is also the same for relationships between $R_{10/10}$ and
$\bar{l}$ in simulation.

\section{Conclusion}

The variation of amino acid frequencies and the variation of genomic
base compositions can be explained in a unified framework based on
genetic code evolution. According to the model, we show that the
genetic code multiplicity plays central roles in explanation of the
variation of amino acid frequencies of contemporary species. The
theoretical results agree with the experimental observations not
only in the variation trends but also in the fine structures or
superfine structures. And we also show that the codon chronology
plays central roles in explanation of the variation of genomic base
compositions of contemporary species. Our results agree with the
experimental observations in many details, such as S-shaped or
inverse S-shaped profiles in Fig. 2, differences of ranges of amino
acid frequencies between Archaea and Eukarya in Fig. 4,
characteristics of step and junction in Fig. 6e, bending direction
in Fig. 8 etc, which confirm the validity of our theory. We also
explained that the mechanism to determine the variation of amino
acid frequencies originated before the stage when three domains
began to branch. So we conclude that the evolution of genetic code
is the initial driving force in molecular evolution, and the
evolution of genetic code in the primordial period determines the
variations of amino acid frequencies and genomic base compositions
of contemporary species.

\section*{Acknowledgments}

We thank Hefeng Wang for valuable discussions. Supported by NSF of
China Grant No. of 10374075.

\begin{figure}
\centering{
\includegraphics[width=160mm]{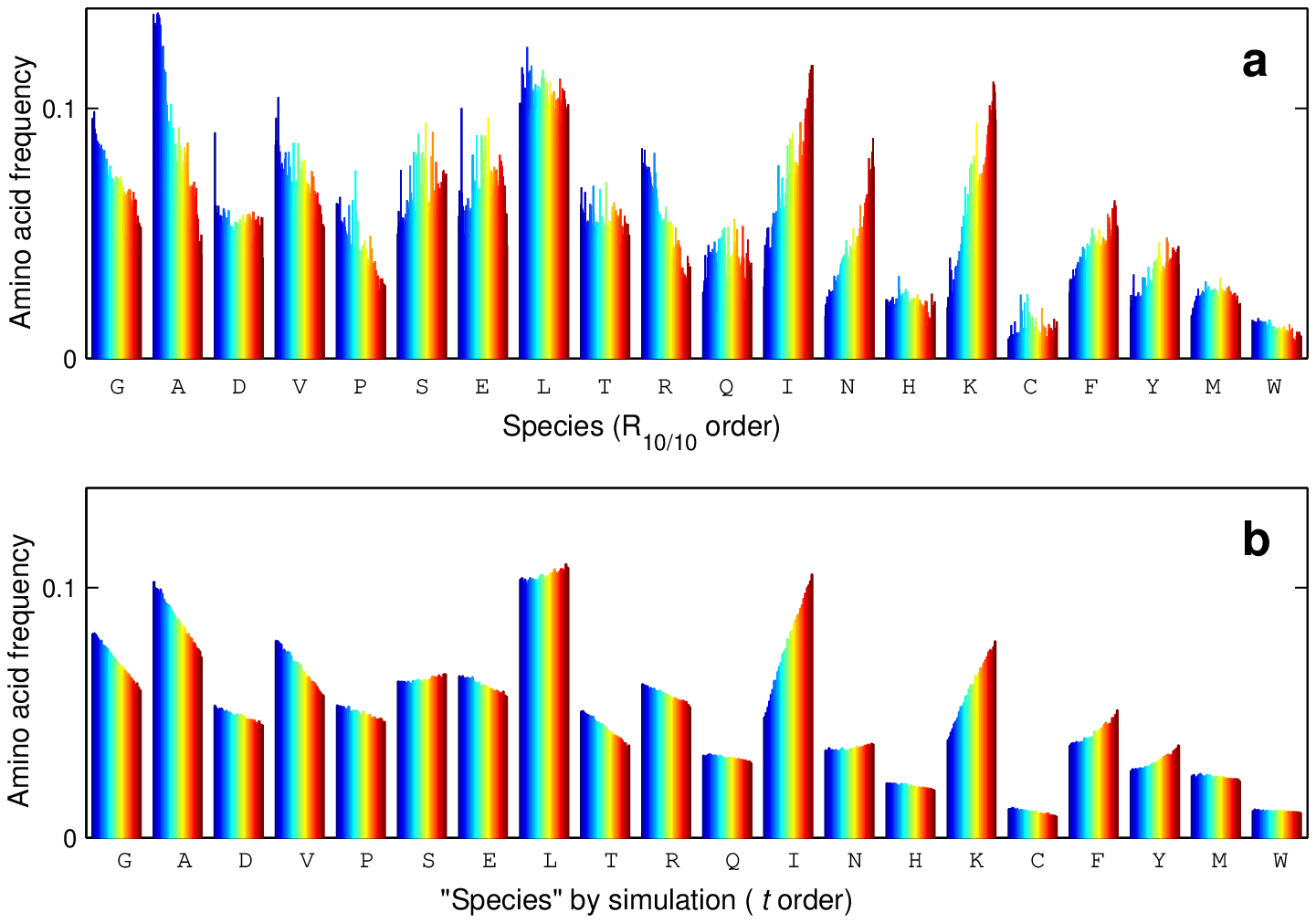}
} \label{fig} \caption{\small {\bf Variation trends of amino acid
frequencies.} {\bf a,} Experimental observations of the variation
trends of amino acid frequencies are based on the data of $106$
species in PEP, which agree with the results based on the data of
$803$ species in NCBI (Fig. S4). Each column represents a variation
trend of one of the $20$ amino acids, where the $106$ species are
aligned from left to right by $R_{10/10}$ order. The $20$ amino
acids are aligned chronologically from left to right. {\bf b,}
Theoretical results of the variation trends of amino acid
frequencies. The $30$ simulated species are aligned by $t$ order.
The variation trends in simulation fit the experimental observations
for each amino acid.}
\end{figure}

\clearpage

\begin{figure}
\centering{
\includegraphics[width=130mm]{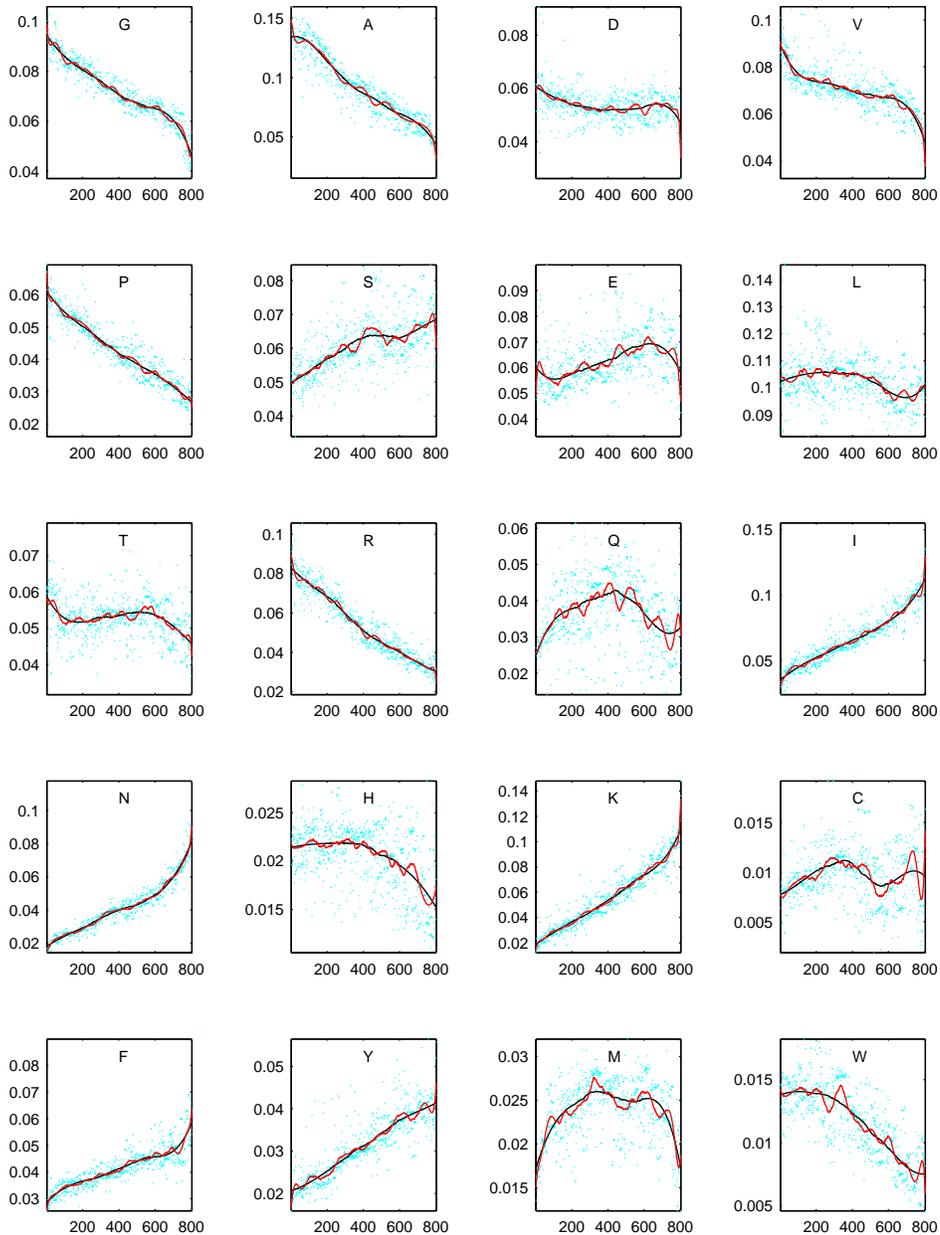}
} \label{fig} \caption{\small {\bf Fine structures and superfine
structures of the variation of amino acid frequencies in
experimental observations.} This is based on the data of $803$
species in NCBI (celeste dots), which are alined by $R_{10/10}$
order for each amino acid. The S-shaped or inverse S-shaped profiles
are the fine-structures (smoothed lines: blank, span=401, degree=3).
The detailed fluctuations represent the superfine-structures
(smoothed lines: red, span=201, degree=7). For each subplot, x-axis
represents the species, y-axis represents amino acid frequencies,
and the smoothed line for the variation of amino acid frequency is
according to Savitzky-Golay method. These conventions are valid for
other similar figures in this paper.}
\end{figure}

\clearpage

\begin{figure}
\centering{
\includegraphics[width=130mm]{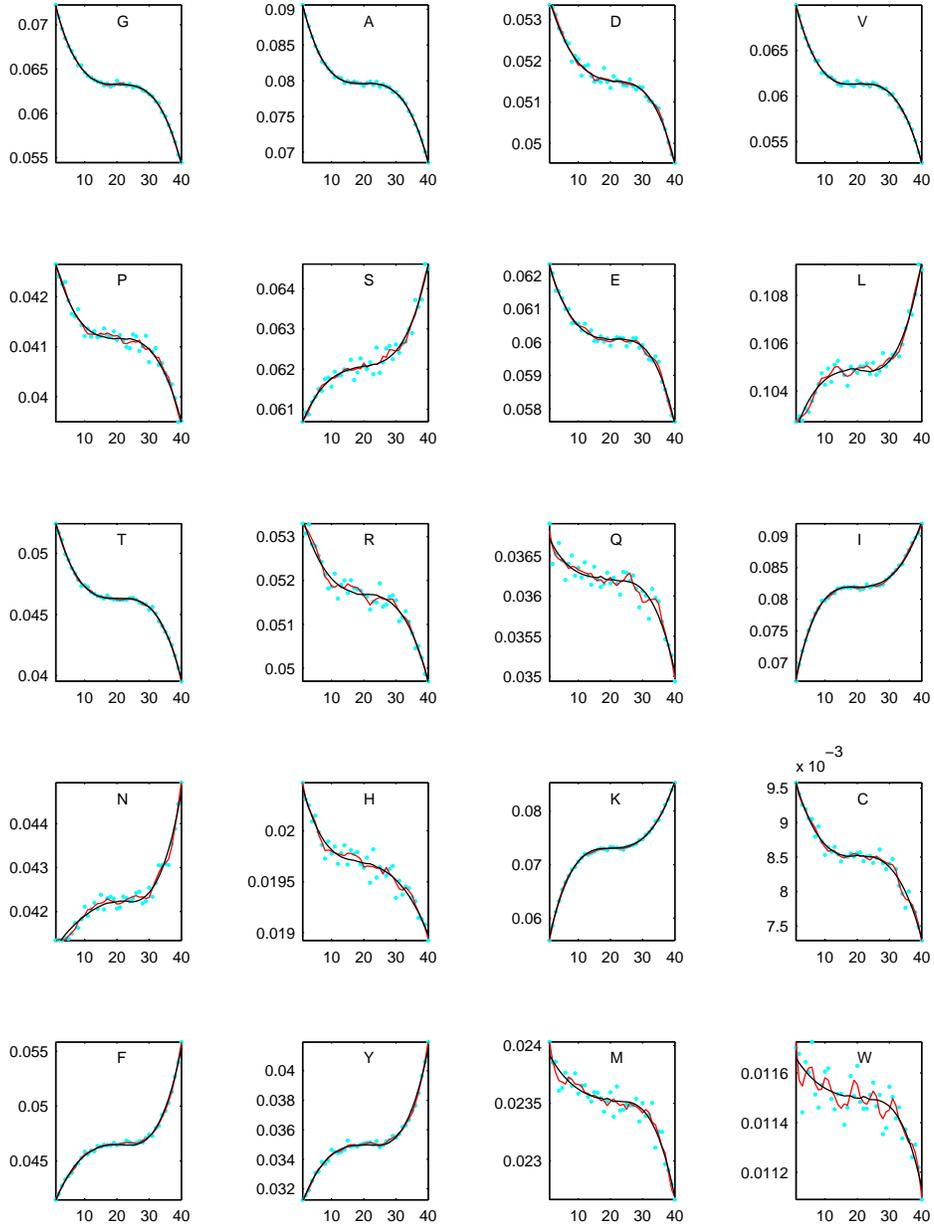}
} \label{fig} \caption{\small {\bf Fine structures and superfine
structures of the variation of amino acid frequencies in theoretical
results.} This is based on the data of $n_m=40$ species (celeste
dots) in simulation. The calculation of amino acid frequencies for
each simulated species is based on $n_p=400,000$ protein sequences
generated by the model. The species are alined by $t$ order in
x-axis for each amino acid. The S-shaped or inverse S-shaped
profiles are the fine-structures (smoothed lines: blank, span=31,
degree=3). The detailed fluctuations represent the
superfine-structures (smoothed lines: red, span=7, degree=3). The
theoretical results agree with experimental observations (Fig. 2) in
general.}
\end{figure}

\clearpage

\begin{figure}
\centering{
\includegraphics[width=120mm]{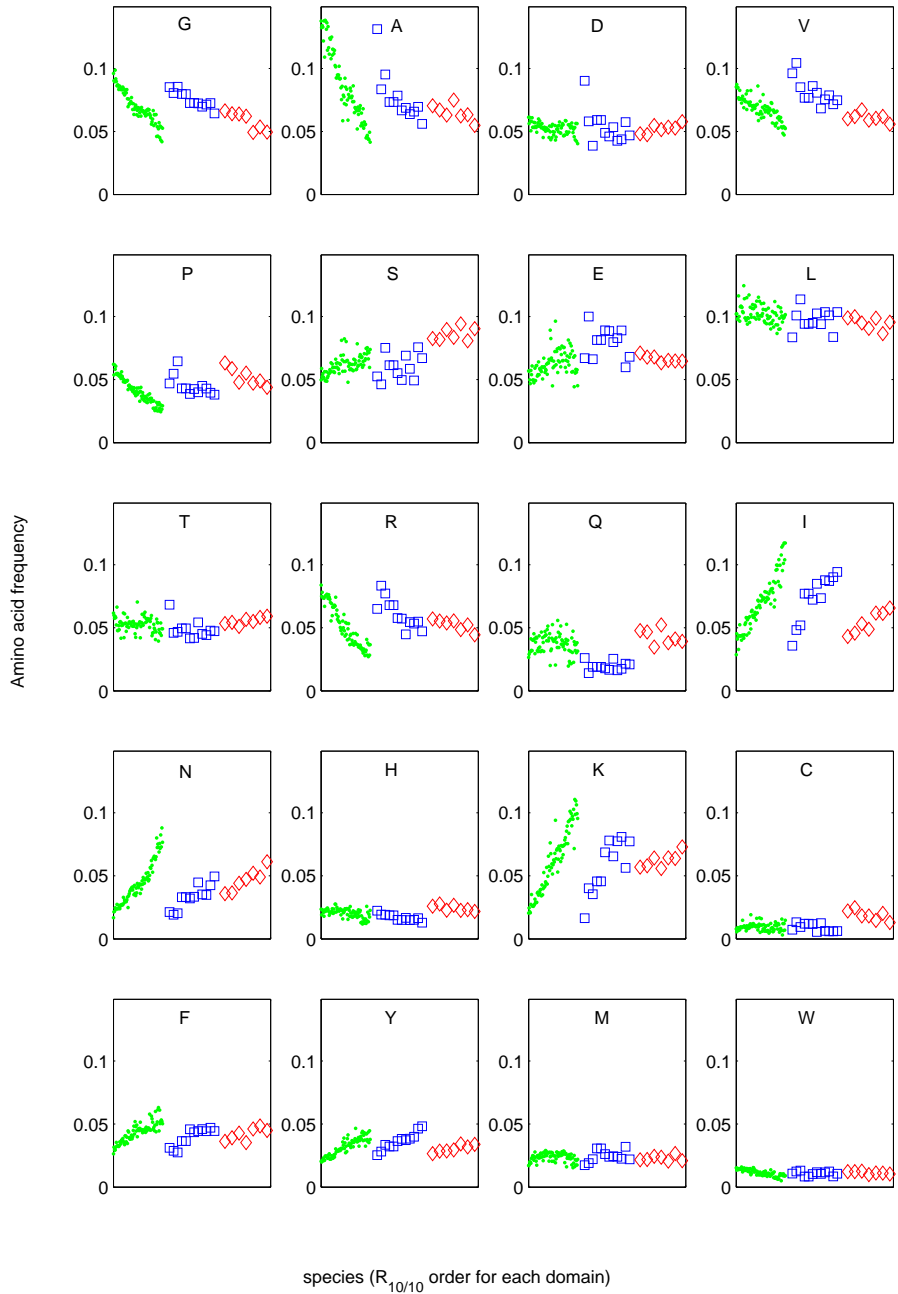}
} \label{fig}  \caption{\small{\bf Variation of amino acid
frequencies for three domains in experimental observations.} The
variation trends of amino acid frequencies are the same in general
for three domains, which are aligned from left to right for each
amino acid (Eubacteria: green dots, Archaebacteria: Blue square, and
Eukaryotes: red diamonds). The amino acid frequencies are about the
same for Archaea and Eubacteria. However, there are obvious
differences of amino acid frequencies between Archaea (Eubacteria)
and eukaryotes (namely, there is an upward shift when increasing or
a downward shift when decreasing for the eukaryotes).}
\end{figure}

\clearpage

\begin{figure}
\centering{
\includegraphics[width=120mm]{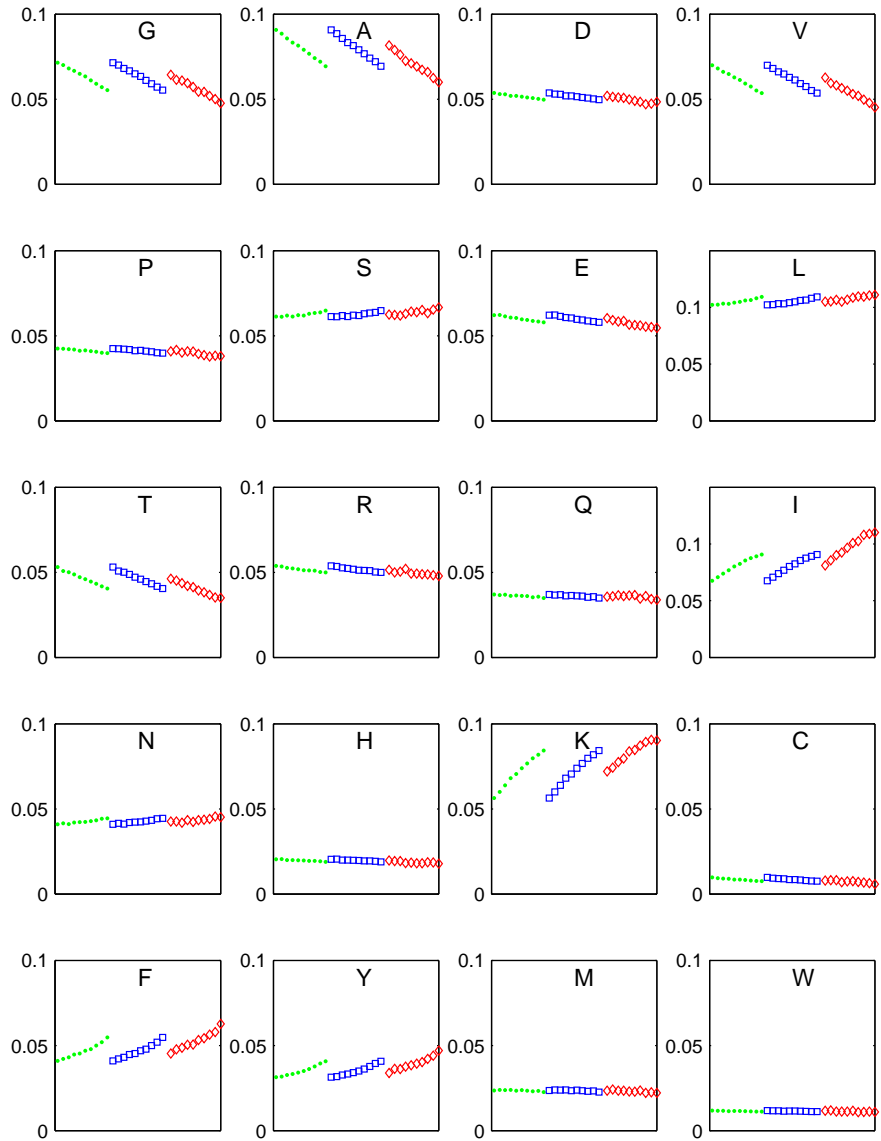}
} \label{fig} \caption{\small{\bf Variation of amino acid
frequencies for three domains in theoretical results.} The
theoretical results agree with the experiment observations (Fig. 4)
in general. The evolutionary trends of amino acid frequencies are
the same for three domains in the simulation. And the amino acid
frequencies are the same for Archaea and Eubacteria; while there are
also upward or downward shifts of amino acid frequencies between
Archaea (Eubacteria) and eukaryotes in the simulation. According to
the phylogeny of three domains, Eukaya appeared latest in the
evolution. Therefore, we input the initial amino acid frequencies
$iAAF_{\mbox{\tiny NCBI}}$ to simulate the evolution of amino acid
frequencies for both Eubacteria (green dots) and Archaea (Blue
squares) firstly. Then, we input $iAAF_{\mbox{\tiny medium}}$ (the
value of amino acid frequencies in medium time in the above
simulate) as initial amino acid frequencies to simulate the
evolution of amino acid frequencies for Eukaryotes (Red diamonds).}
\end{figure}

\clearpage

\begin{figure}
\centering{
\includegraphics[width=40mm]{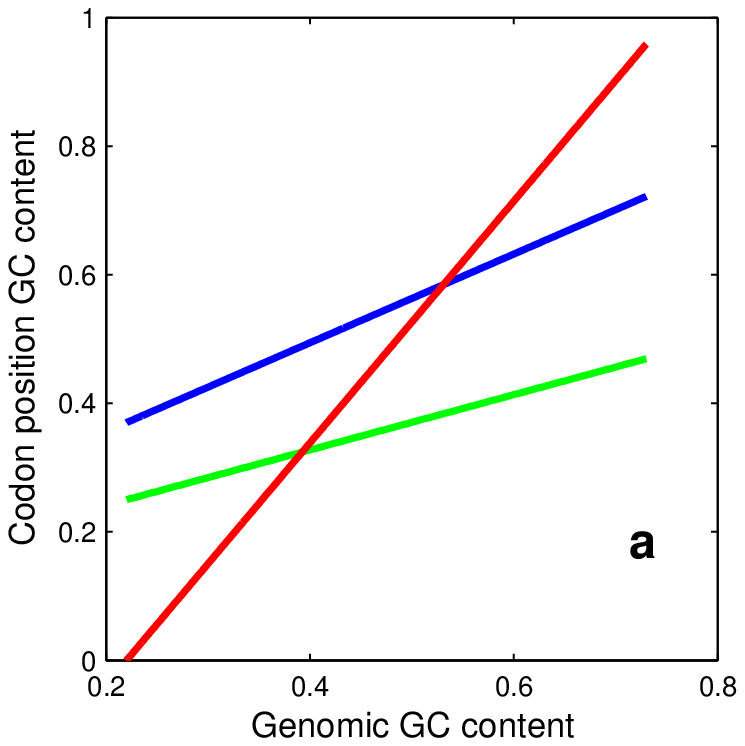}
\includegraphics[width=40mm]{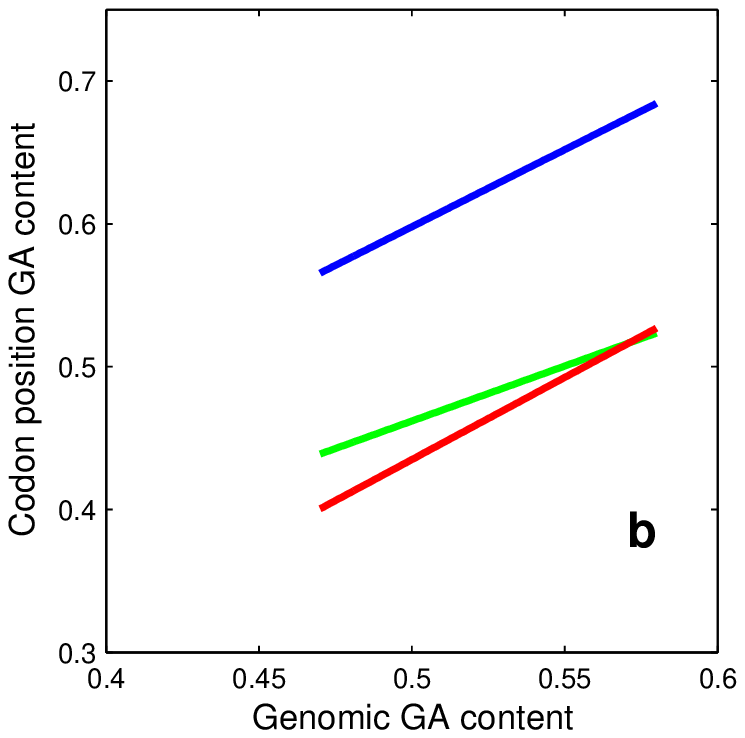}
\includegraphics[width=40mm]{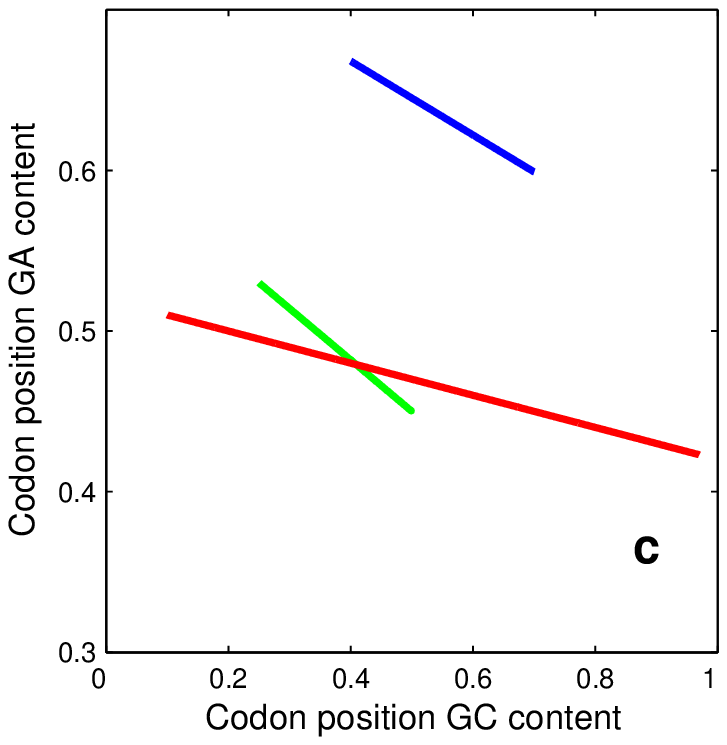}
\includegraphics[width=40mm]{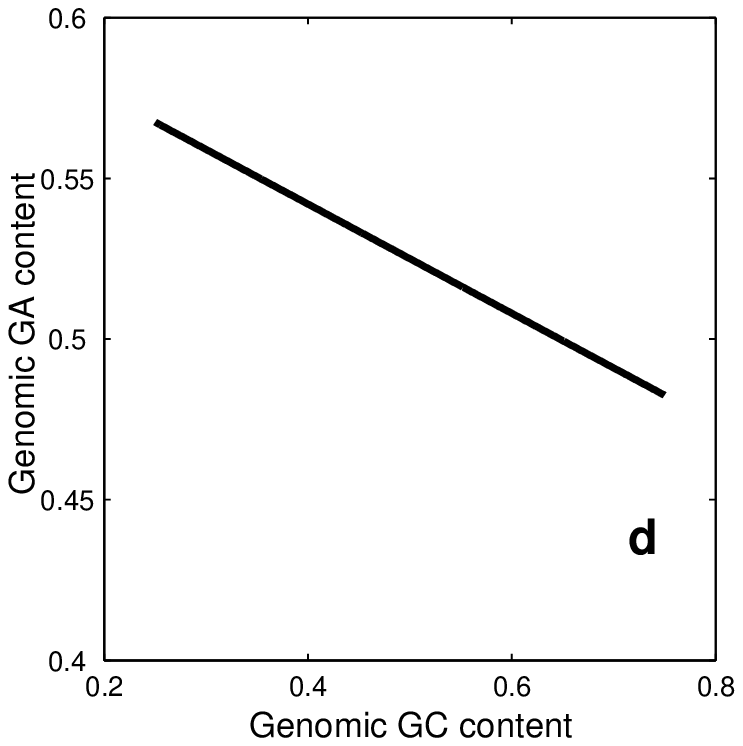}\\
\includegraphics[width=40mm]{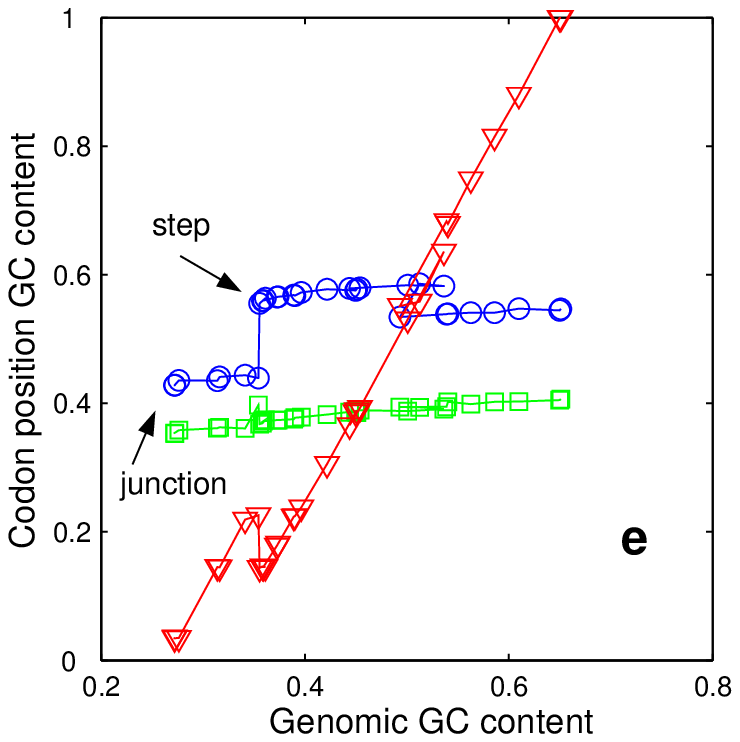}
\includegraphics[width=40mm]{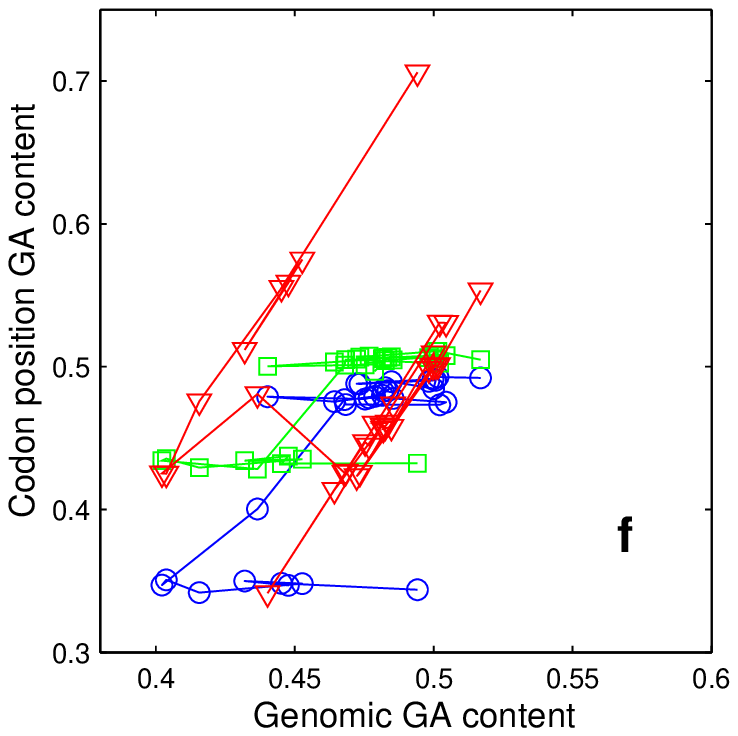}
\includegraphics[width=40mm]{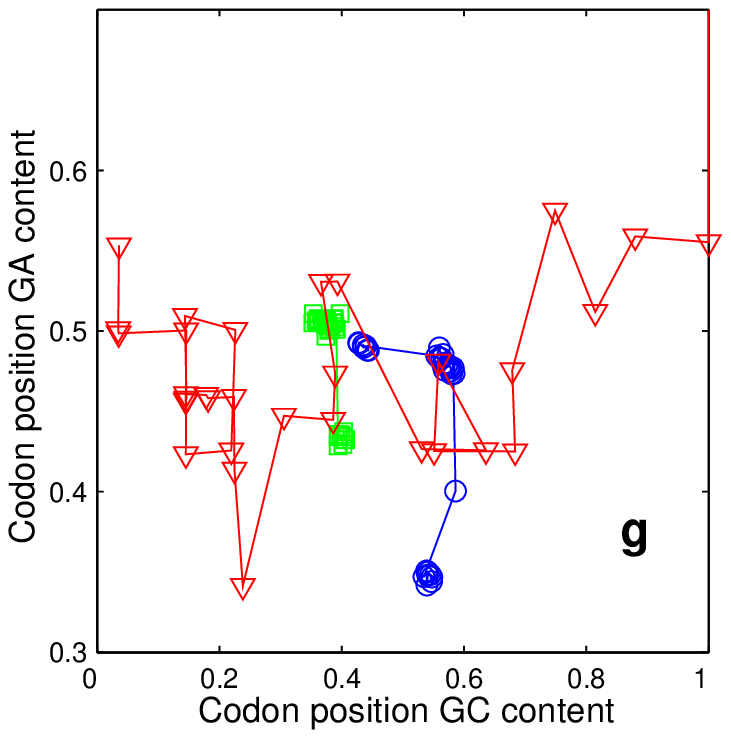}
\includegraphics[width=40mm]{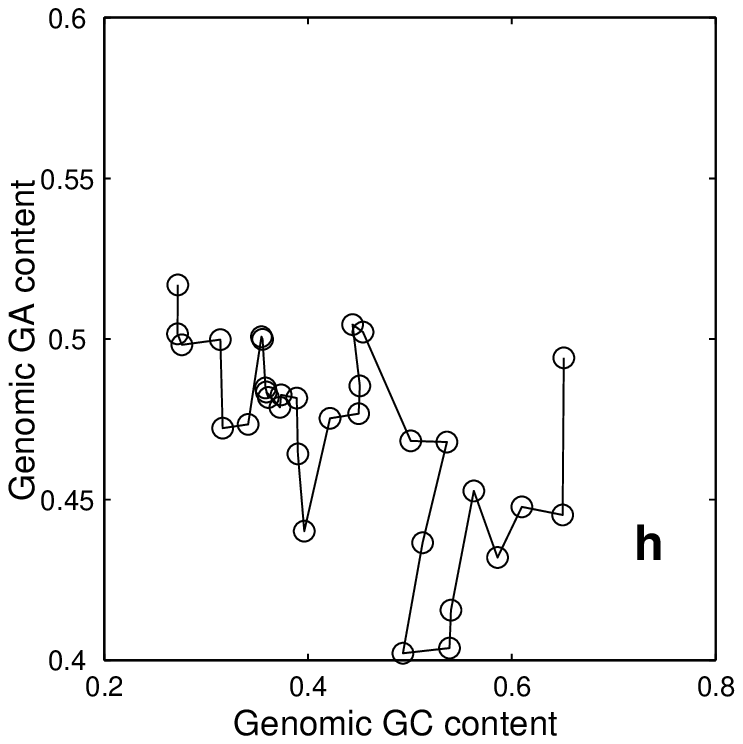}\\
\includegraphics[width=40mm]{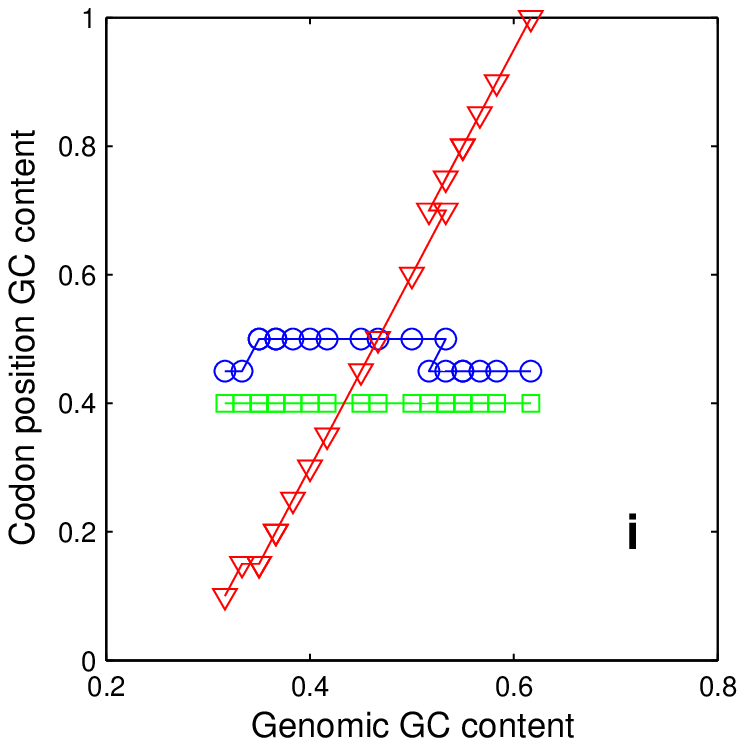}
\includegraphics[width=40mm]{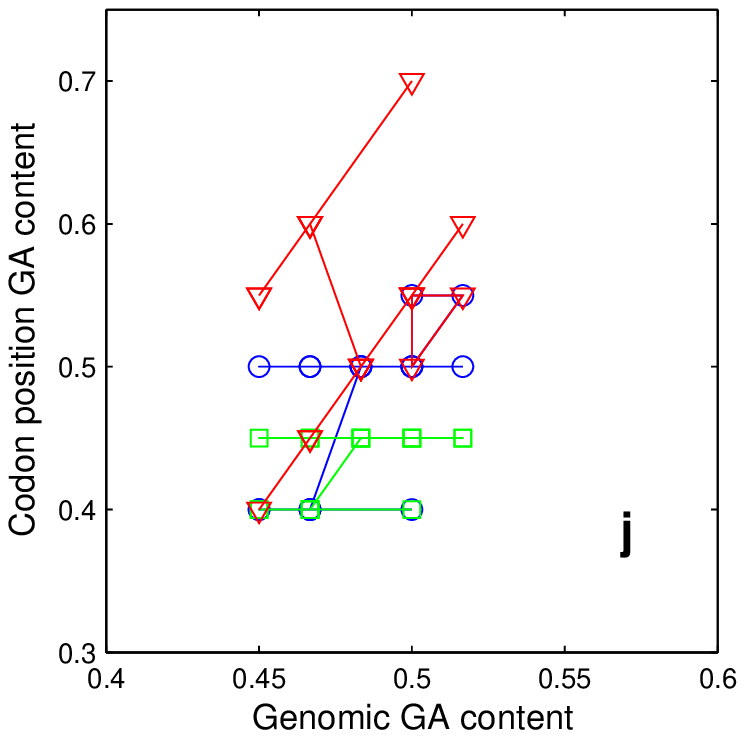}
\includegraphics[width=40mm]{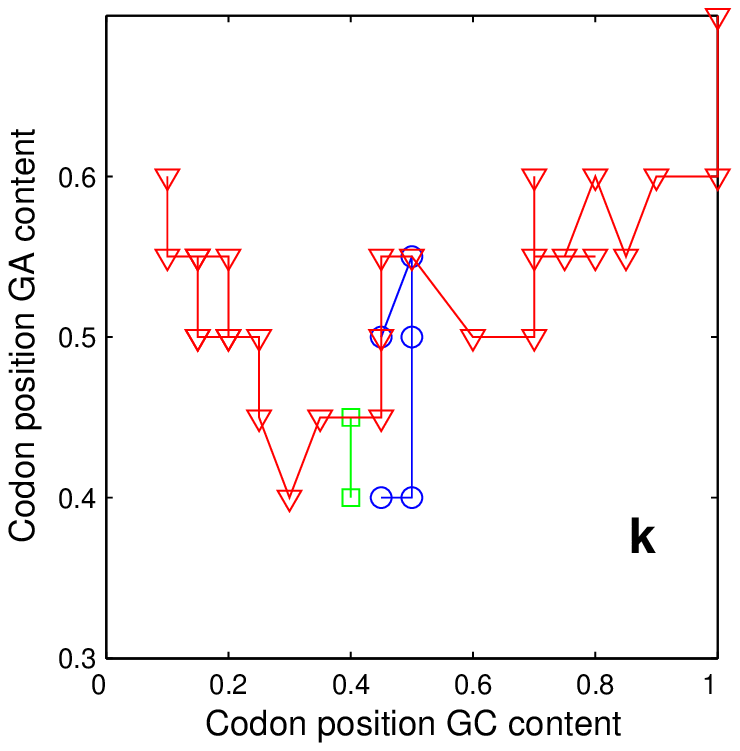}
\includegraphics[width=40mm]{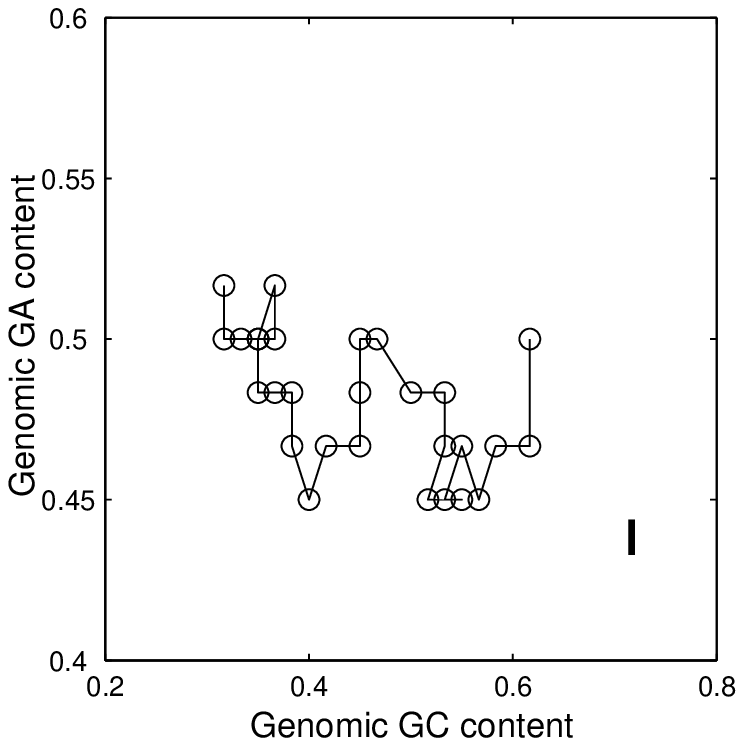}}
\label{fig} \caption{\small {\bf Correlations of genomic GC content
or GA content and codon position GC content or GA content.} The
results for 1st, 2nd, and 3rd codon positions are represent by blue
circles, green squares and red triangles respectively. {\bf a}-{\bf
d,} Correlations of base compositions in codon positions and genomic
base compositions in experimental observations; {\bf e}-{\bf h,}
Correlations of base compositions in codon positions and genomic
base compositions in theoretical results. Simulations by the model
are based on genetic code multiplicity, which agree with the above
experimental observations in either evolutionary trends or some
detailed characters; {\bf i}-{\bf l,} Correlations of number of
bases in codon positions and total number of bases are based on Tab.
6, which mainly determines the results in corresponding
simulations.}
\end{figure}

\clearpage

\begin{figure}
\centering{
\includegraphics[width=60mm]{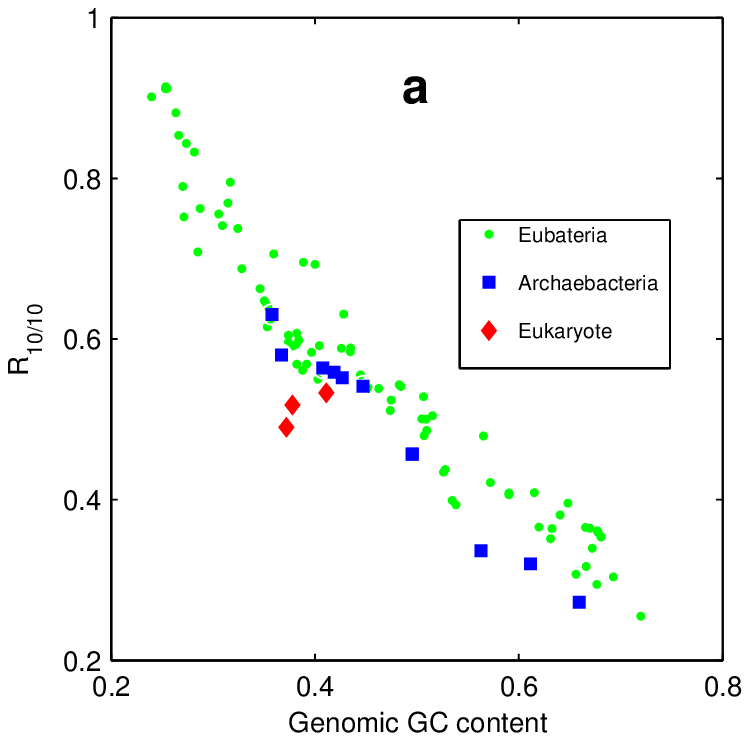}
\includegraphics[width=63mm]{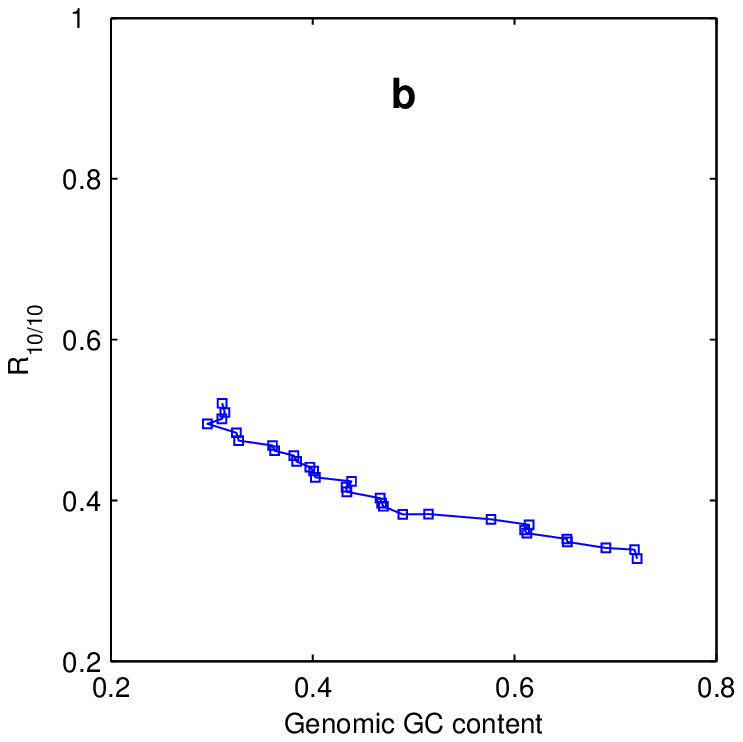}
} \label{fig} \caption{\small {\bf The relationship between genomic
GC content and the variation of amino acid frequencies.} {\bf a,}
The GC content declines with the ration $R_{10/10}$ according to the
biological data. {\bf b,} The simulation agrees with the
experimental observation in the variation trend between GC content
and $R_{10/10}$.}
\end{figure}

\clearpage

\begin{figure}
\centering{
\includegraphics[width=120mm]{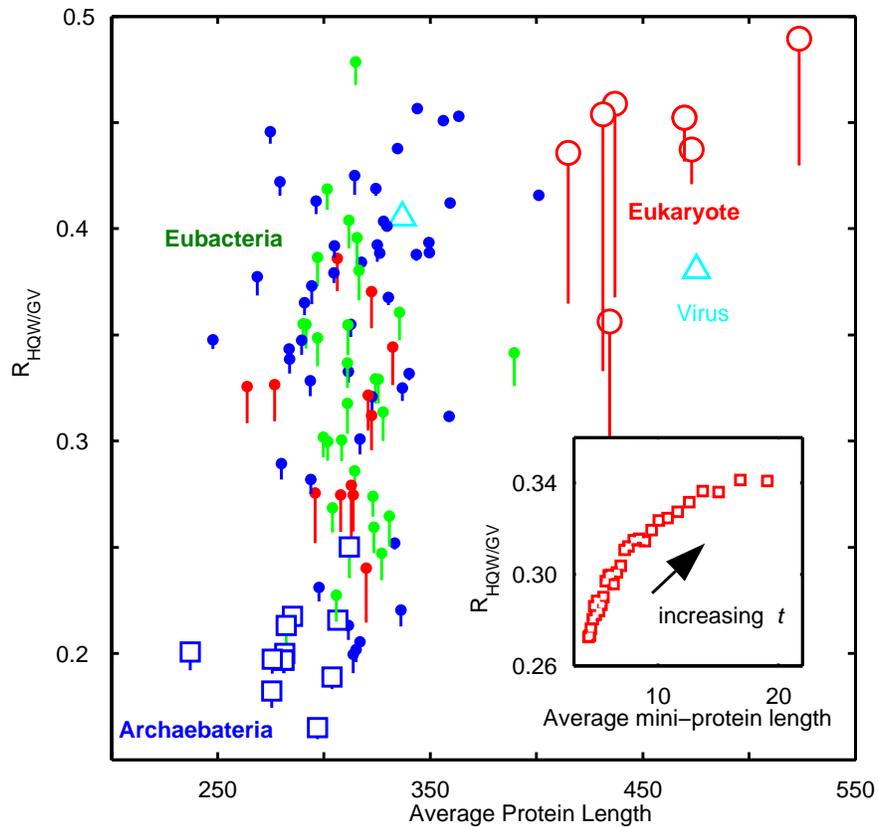}
} \label{fig} \caption{\small {\bf The relationship between the
average protein length and the ratio $R_{HQW/GV}$.} The species in
three domains cluster together in three areas respectively. The
genome sizes of species is represented by tails below the
corresponding dots (larger genome size: long red tail; medium genome
size: medium green tail and small genome size: short blue tail).
{\bf Embedded,} The simulation of the relationship between average
protein length and the ratio $R_{HQW/GV}$, especially the bedding
direction, agrees with the experimental observation.}
\end{figure}

\clearpage

\begin{center}
{\Large SUPPLEMENTARY FIGURES AND TABLES}
\end{center}

\section*{Guide to figures and tables}

\noindent{\bf Group I:} Fig. 1, Fig. S1-S4, Tab. 1. Variation trends
of amino acid frequencies.

\noindent{\bf Group II:} Fig. 2-3, Fig. S5-S8. Fine structures and
superfine structures of the variation of amino acid frequencies.

\noindent{\bf Group III:} Fig. 4-5, Fig. S9. Variation of amino acid
frequencies for three domains.

\noindent{\bf Group IV:} Fig. 6, Fig. S10. Variation of genomic base
compositions.

\noindent{\bf Group V:} Fig. 7-8. Relationship between amino acid
frequencies and genomic base compositions, and relationship between
amino acid frequencies and average protein length

\noindent{\bf Group VI:} Fig. S11-S14, Tab. 2-6. The model based on
genetic code multiplicity and codon chronology.

\clearpage

\begin{center}
{\Large SUPPLEMENTARY FIGURES S1-S14}
\end{center}

\clearpage

\begin{figure}
\centering{
\includegraphics[width=135mm]{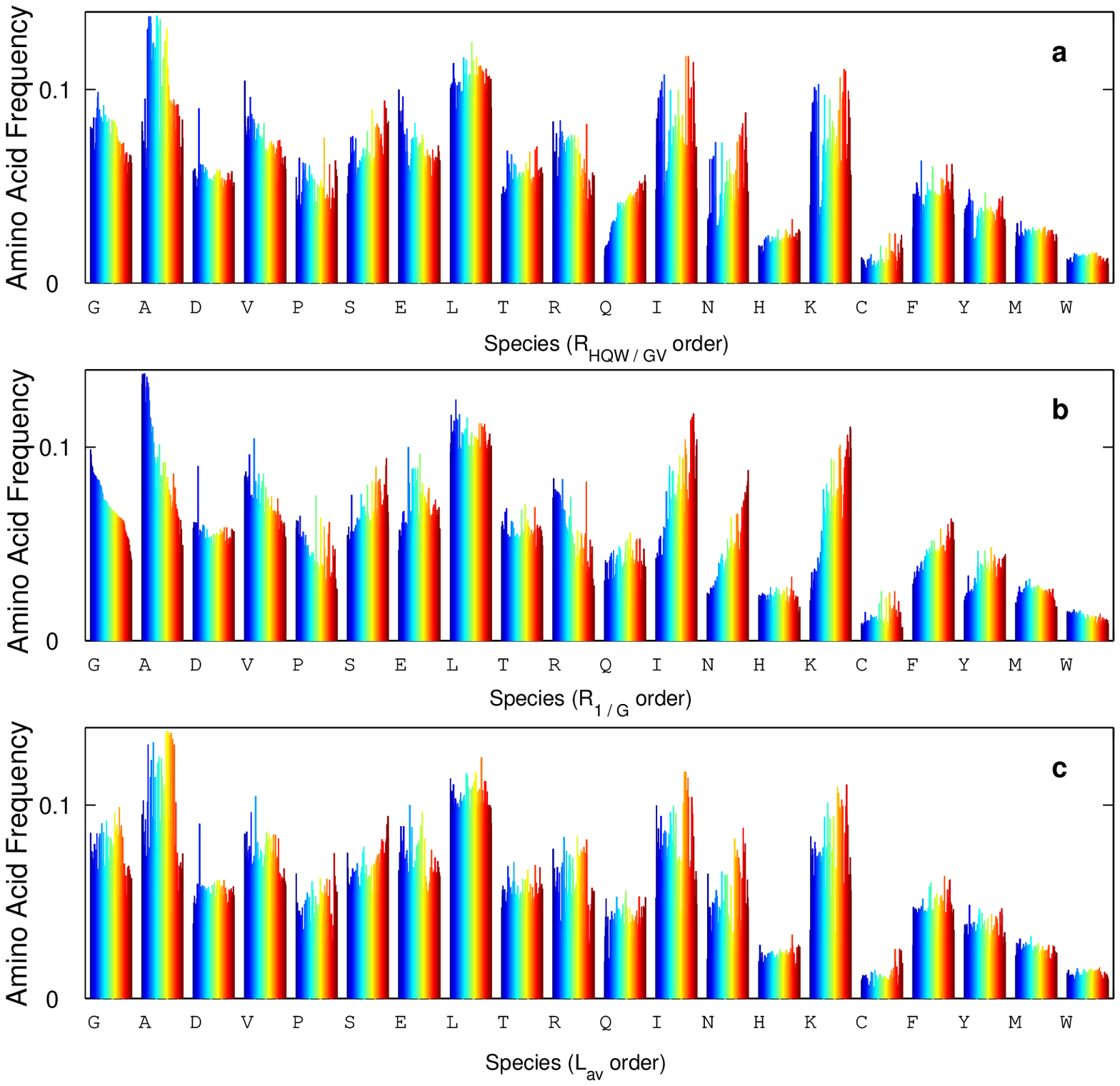}
} \label{fig} \caption{\small {\bf Fig. S1 Variation of amino acid
frequencies.} The evolutionary trends are generally the same for the
Late-early Ratio Orders. {\bf a,} The $R_{HQW/GV}$ order; {\bf b,}
The $R_{1/G}$ order; {\bf c,} The $L_{av}$ order; }
\end{figure}

\clearpage

\begin{figure}
\centering{
\includegraphics[width=120mm]{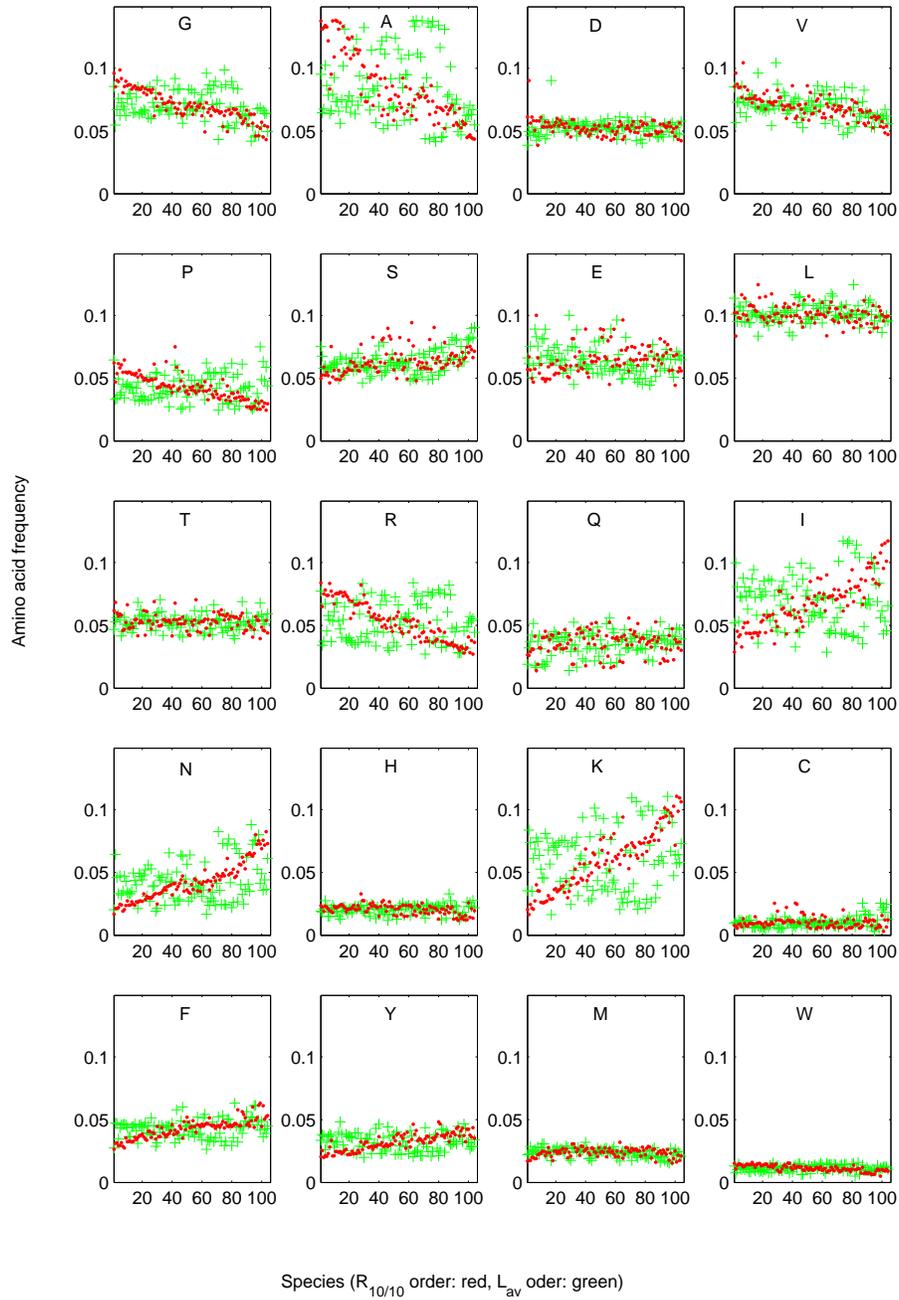}
} \label{fig} \caption{\small {\bf Fig. S2 Comparison of variation
of amino acid frequencies for $R_{10/10}$ order and $L_{av}$ order.}
The evolutionary trends are generally the same for $R_{10/10}$ order
and $L_{av}$ order.}
\end{figure}

\clearpage

\begin{figure}
\centering{
\includegraphics[width=120mm]{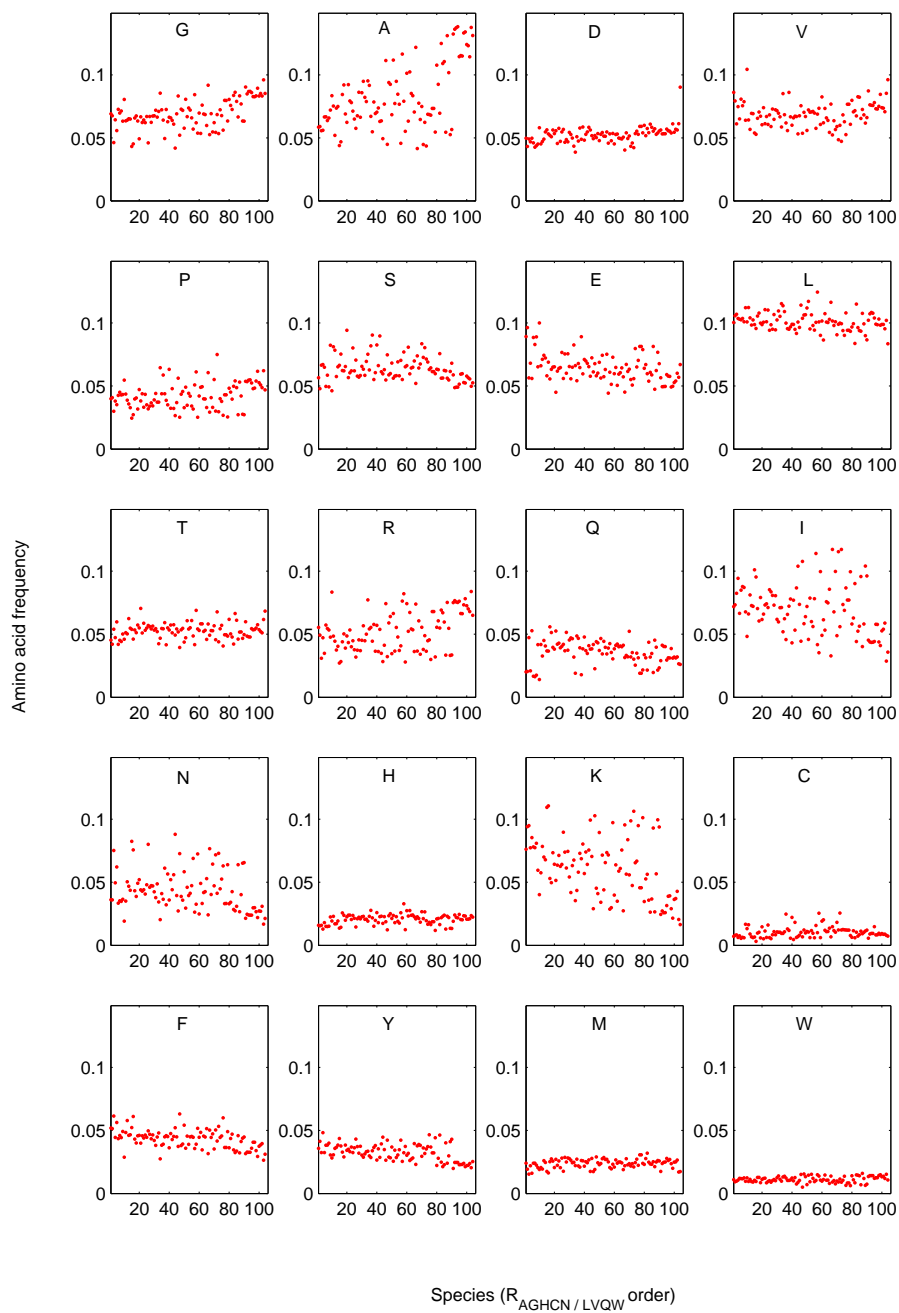}
} \label{fig} \caption{\small {\bf Fig. S3 Variation of amino acid
frequencies for the $R_{AGHCN/LVQW}$ order.} The variation is random
for the Random Ratio Orders.}
\end{figure}

\clearpage

\begin{figure}
\centering{
\includegraphics[width=120mm]{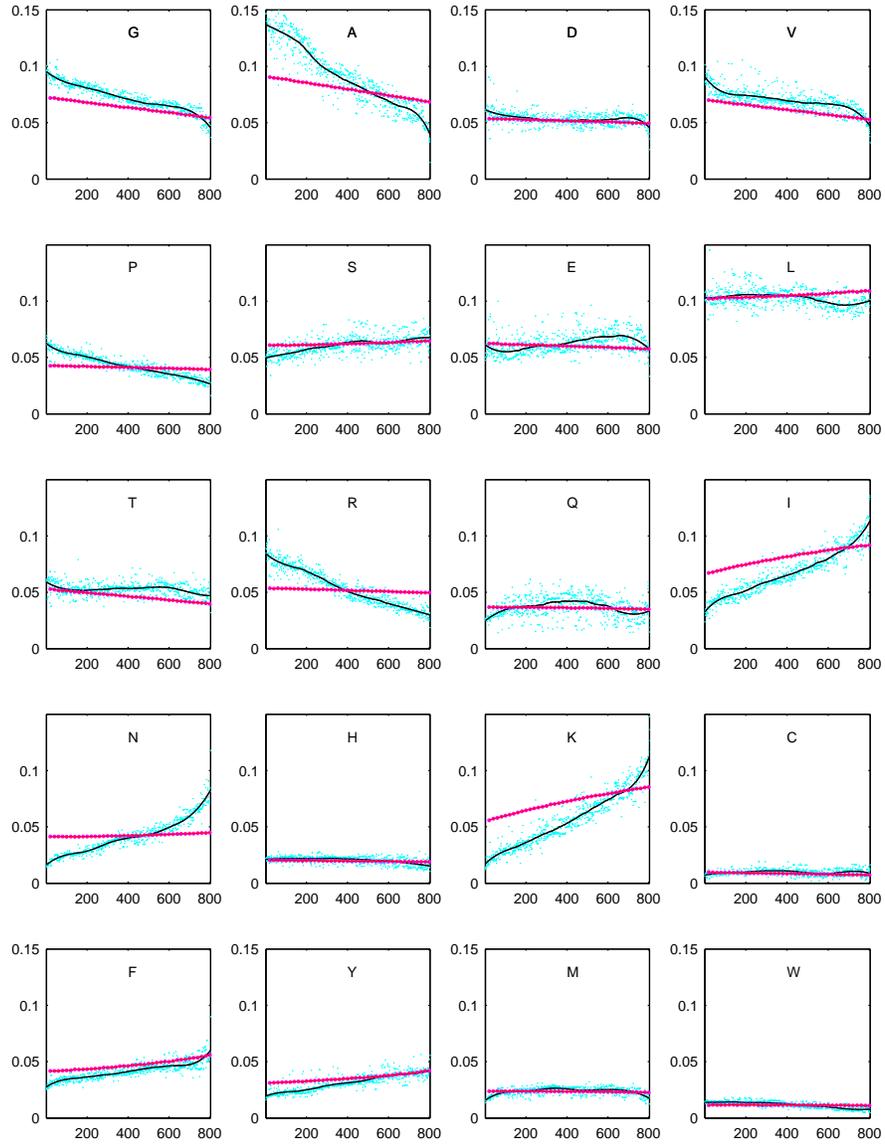}
} \label{fig} \caption{\small {\bf Fig. S4 Variation of amino acid
frquencies.} The amino acid frquencies in experimental observation
are based on the data of $803$ species in NCBI (dots: celeste). The
smoothed line for each amino acid is according to Savitzky-Golay
method (smoothed lines: blank, span=301, degree=3). Simulation of
the variation of amino acid frequencies (dots and smoothed lines:
red, $n_m=40$, $n_p=400,000$, $t=t_1(N_s)$) agrees with the
experimental observations in variation trends.}
\end{figure}

\clearpage

\begin{figure}
\centering{
\includegraphics[width=120mm]{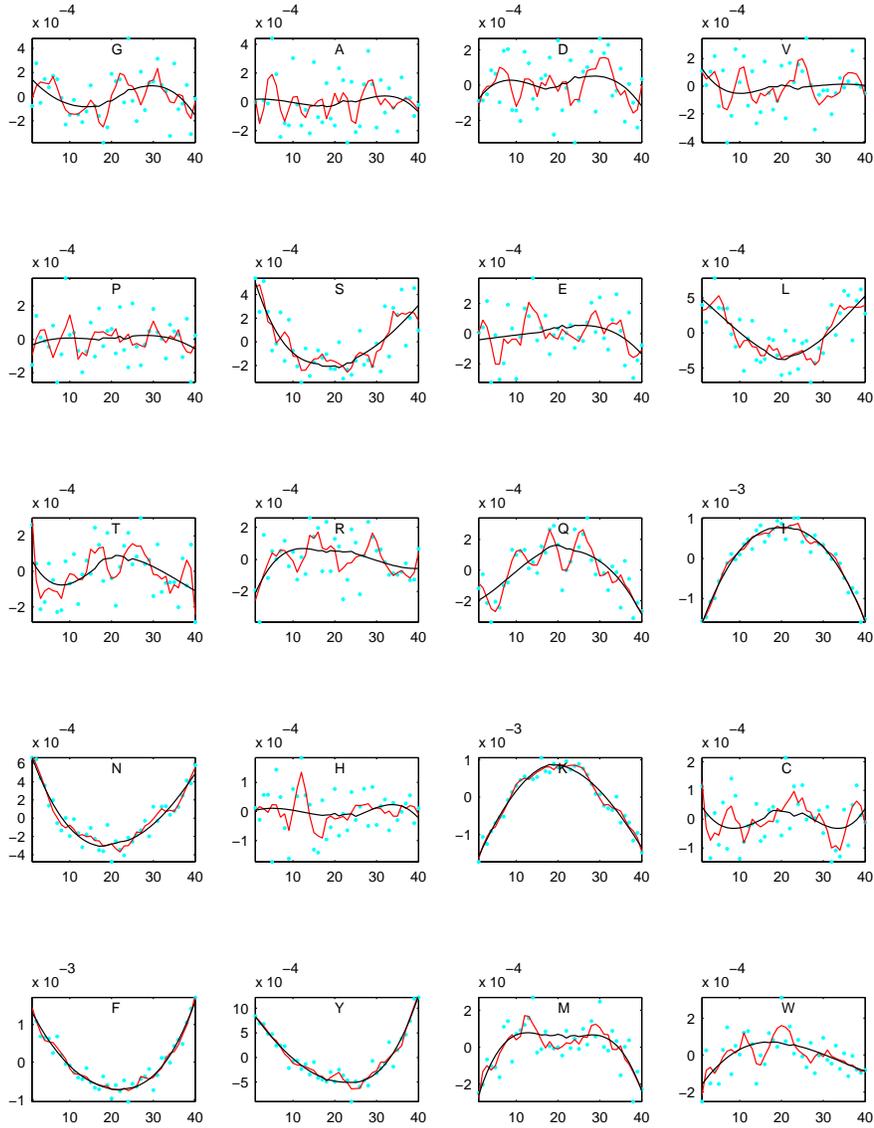}
} \label{fig} \caption{\small {\bf Fig. S5 Superfine structure of
variation of amino acid frquencies in theoretical results.} Firstly,
we obtained amino acid frequencies by simulation ($n_m=40$,
$n_p=400,000$, $t=t_1(N_s)$). Secondly, we obtained the fit line of
variation of amino acid frequencies by least squares for each amino
acid. At last, we obtained the results, i.e., the differences
between the amino acid frequencies and corresponding values of the
fit line by least squares for each amino acid (celeste dots in this
figure. Blank smoothed lines: span=31, degree=3; red smoothed lines:
span=7, degree=3 ).}
\end{figure}

\clearpage

\begin{figure}
\centering{
\includegraphics[width=120mm]{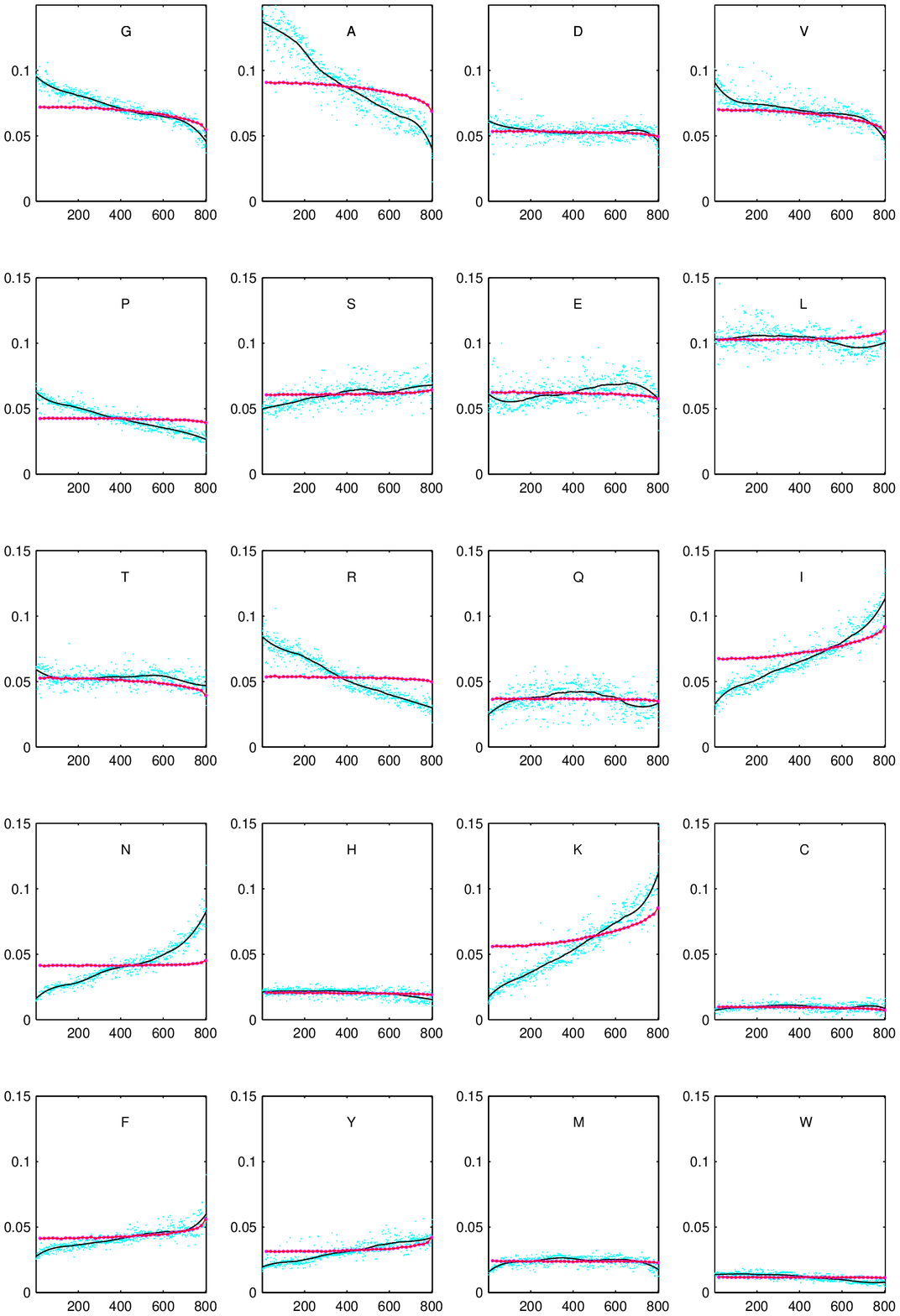}
} \label{fig} \caption{\small {\bf Fig. S6 Simulation of the fine
structure of variation of amino acid frquencies ($t=t_2(N_s)$).}
Simulation of the fine structure of variation of amino acid
frequencies (dots and lines: red, $n_m=40$, $n_p=200,000$), where we
choose $t=t_2(N_s)$, agrees with the experimental observations
(dots: celeste, lines: blank) in bending directions only at right
ends.}
\end{figure}

\clearpage

\begin{figure}
\centering{
\includegraphics[width=120mm]{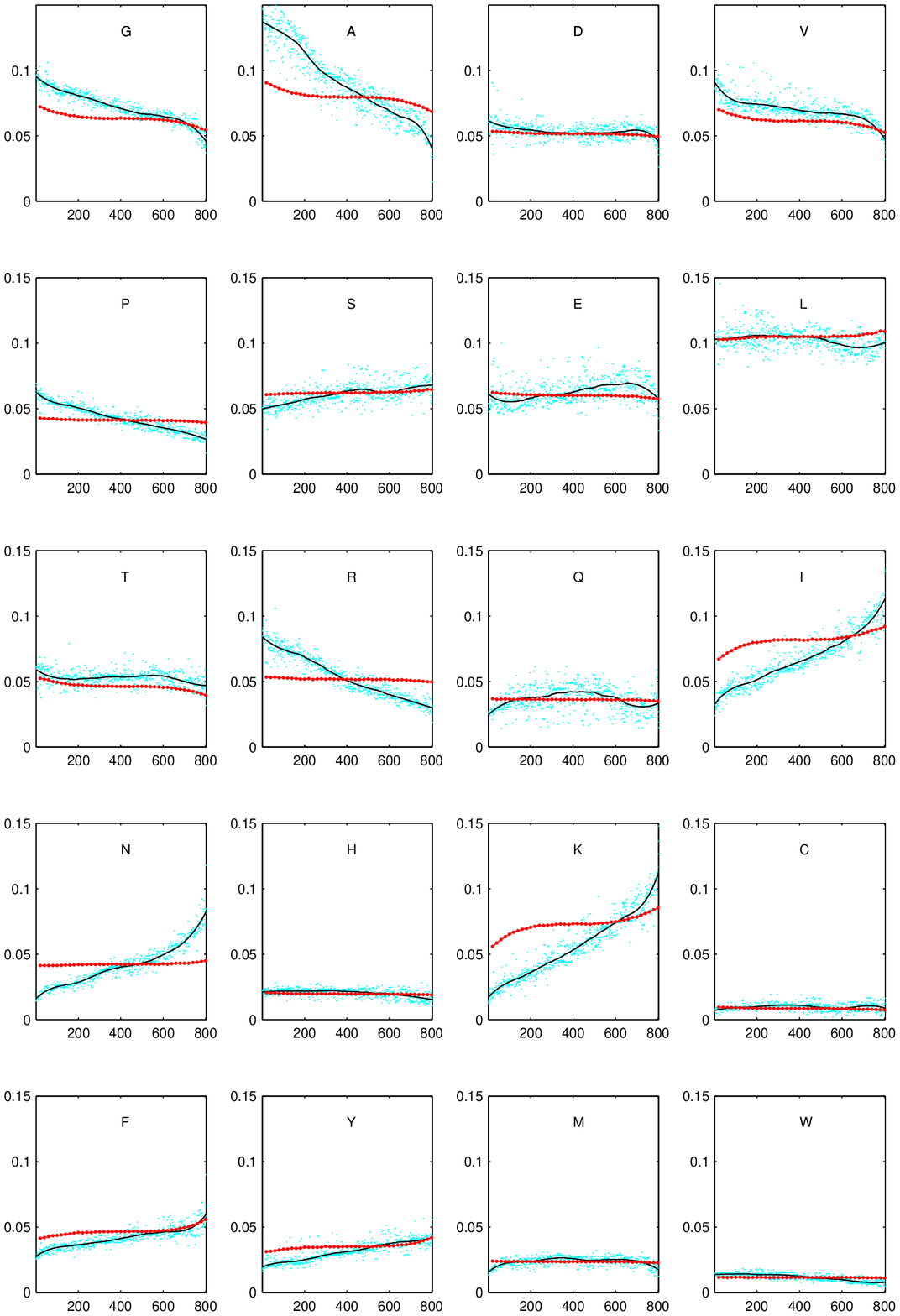}
} \label{fig} \caption{\small {\bf Fig. S7 Simulation of the fine
structure of variation of amino acid frquencies ($t=t_3(N_s)$).}
Simulation of the fine structure of variation of amino acid
frequencies (dots and lines: red, $n_m=40$, $n_p=400,000$), where we
choose $t=t_3(N_s)$, agrees with the experimental observations
(dots: celeste, lines: blank) in bending directions at both right
and left ends.}
\end{figure}

\clearpage

\begin{figure}
\centering{
\includegraphics[width=140mm]{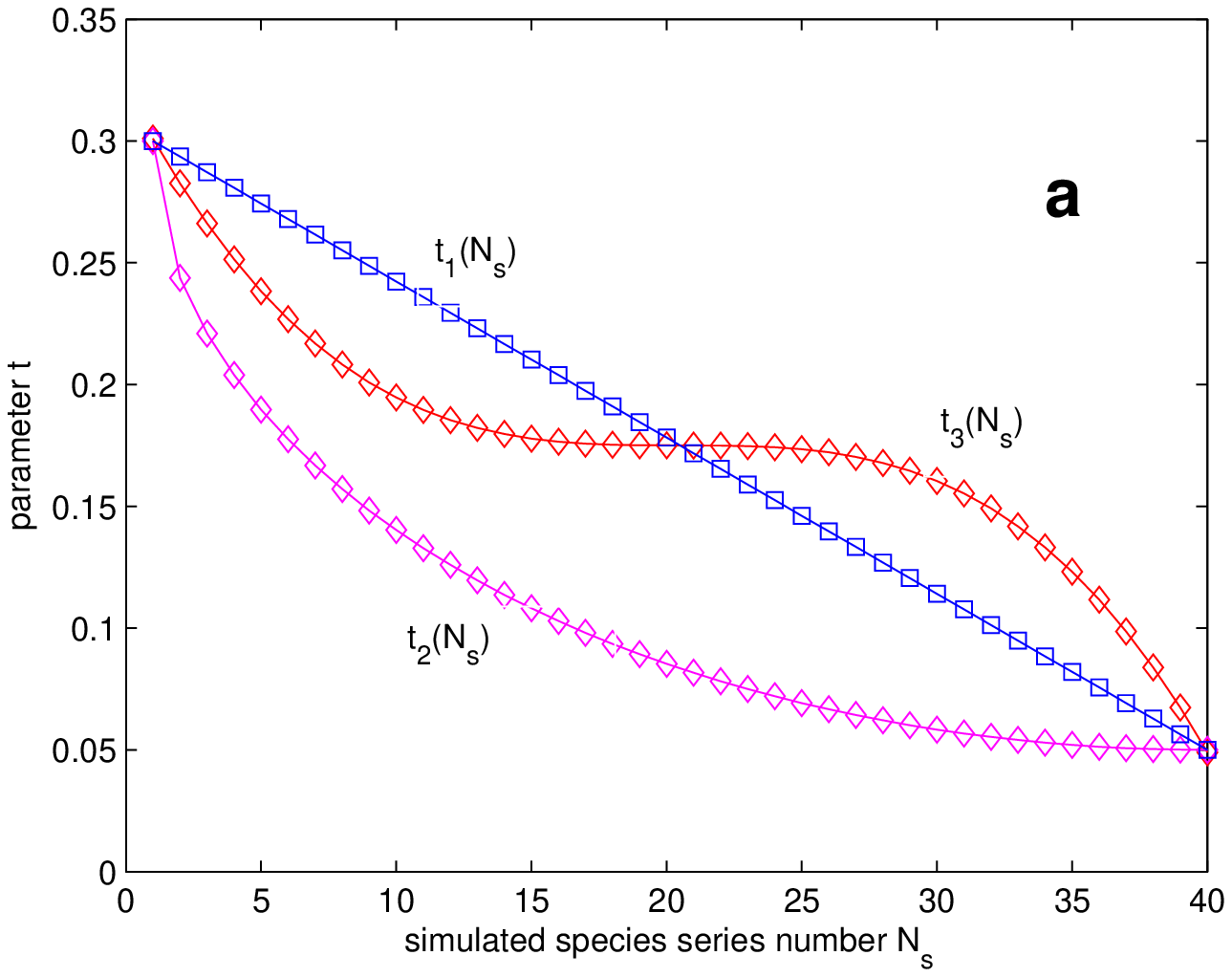}
\includegraphics[width=80mm]{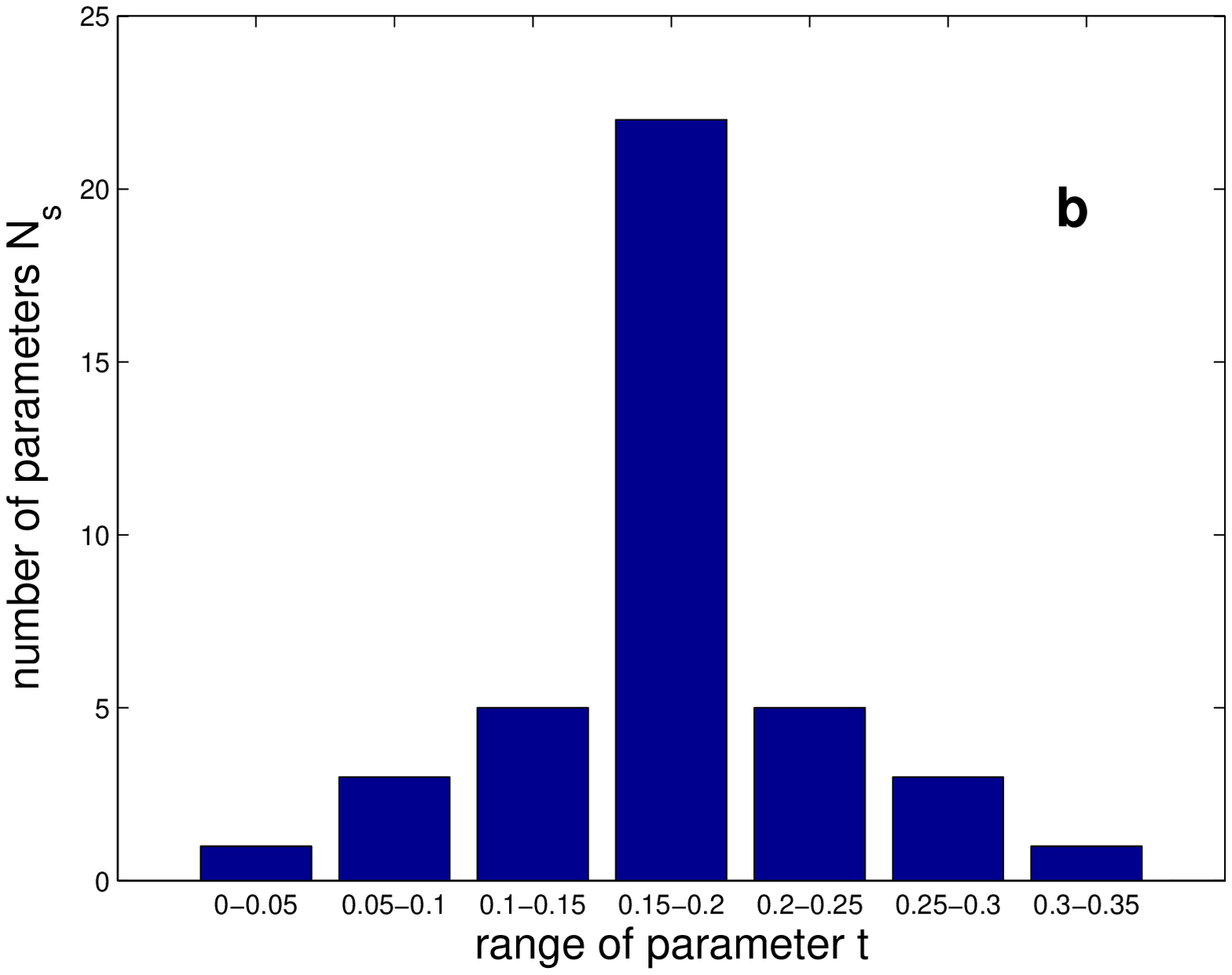}
\includegraphics[width=80mm]{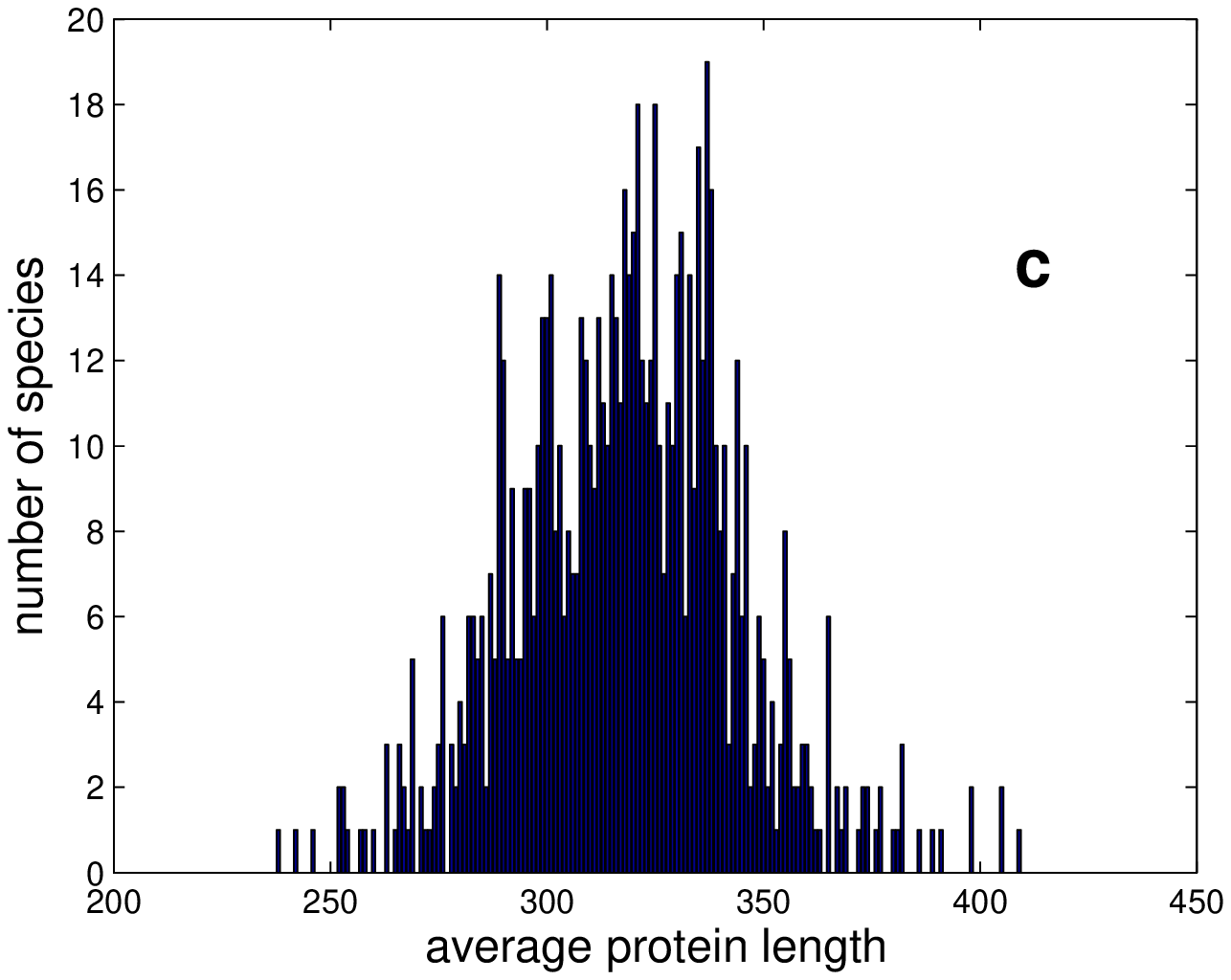}}
\label{fig} \caption{\small {\bf Fig. S8  Explanation of the fine
structure of the variation of amino acid frequencies.} {\bf a,} The
functions between $t$ and $N_s$. The range of parameter $t$ is from
$t_{min}=0.05$ to $t_{max}=0.3$. The series numbers for the
simulated species are $N_s=1,2,...,n_m, n_m=40$. The functions for
the three curves are as follows respectively: (1)
$t_1(N_s)=t_{max}-(N_s-1)(t_{max}-t_{min})/(n_m-1)$, (2)
$t_2(N_s)=t_{max}-(t_{max}-t_{min})\sqrt{1-(N_s-n_m)^2/(n_m-1)^2}$
and (3) $t_3(N_s)=(t_{max}+t_{min})/2-0.000017(N_s-(n_m+1)/2)^3$.
{\bf b,} The distribution of numbers of $N_s$ with respect to the
parameter $t$. {\bf c,} The distribution of $803$ species in NCBI
with respect to average protein length. }
\end{figure}

\clearpage

\begin{figure}
\centering{
\includegraphics[width=120mm]{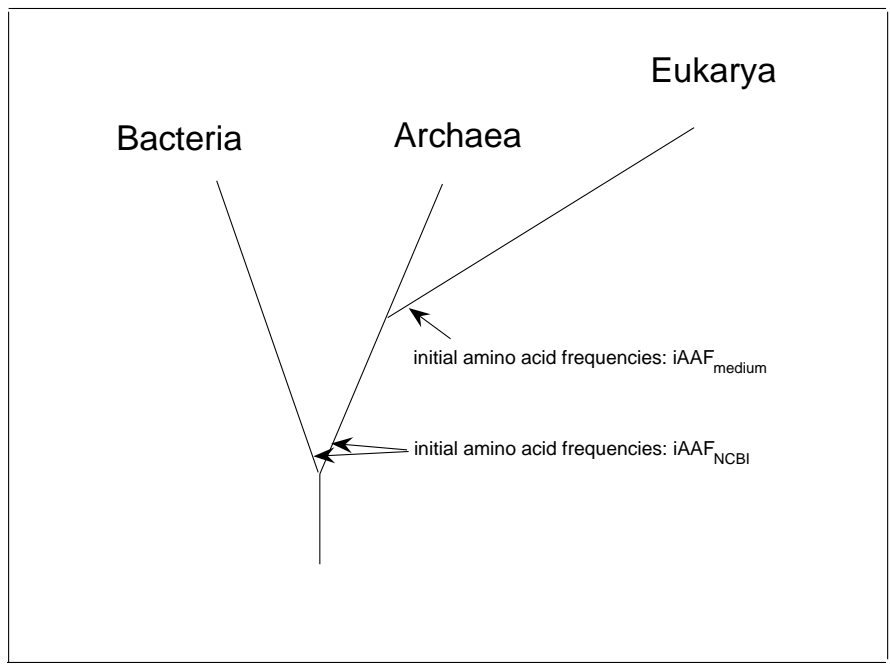}
} \label{fig} \caption{\small {\bf Fig. S9 The phylogeny of three
domains.} In the simulations of variation of amino acid frequencies,
we input $iAAF_{\mbox{\tiny NCBI}}$ as initial amino acid
frequencies for Bacteria and Archaea, and input $iAAF_{\mbox{\tiny
medium}}$ as initial amino acid frequencies for Eukarya.}
\end{figure}

\clearpage

\begin{figure}
\centering{
\includegraphics[width=40mm]{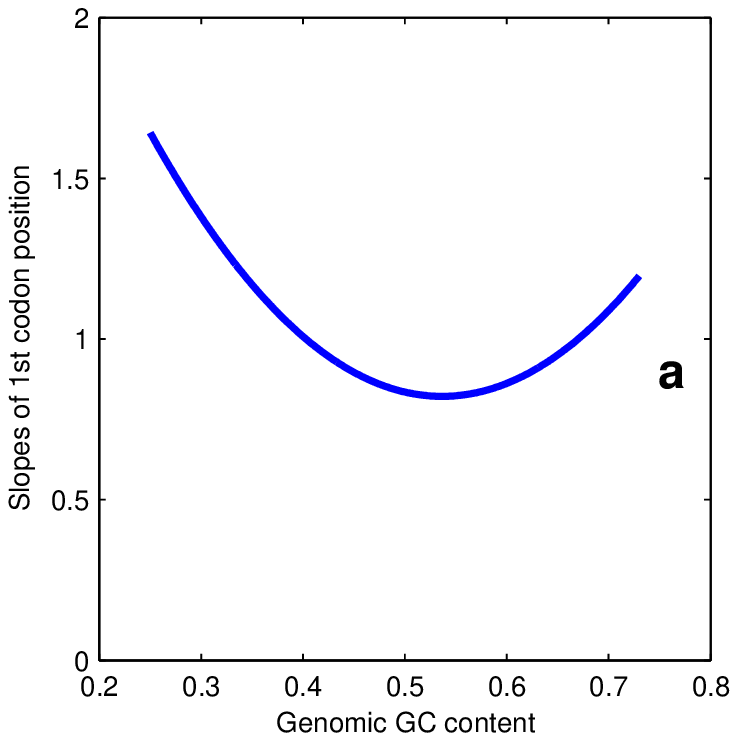}
\includegraphics[width=40mm]{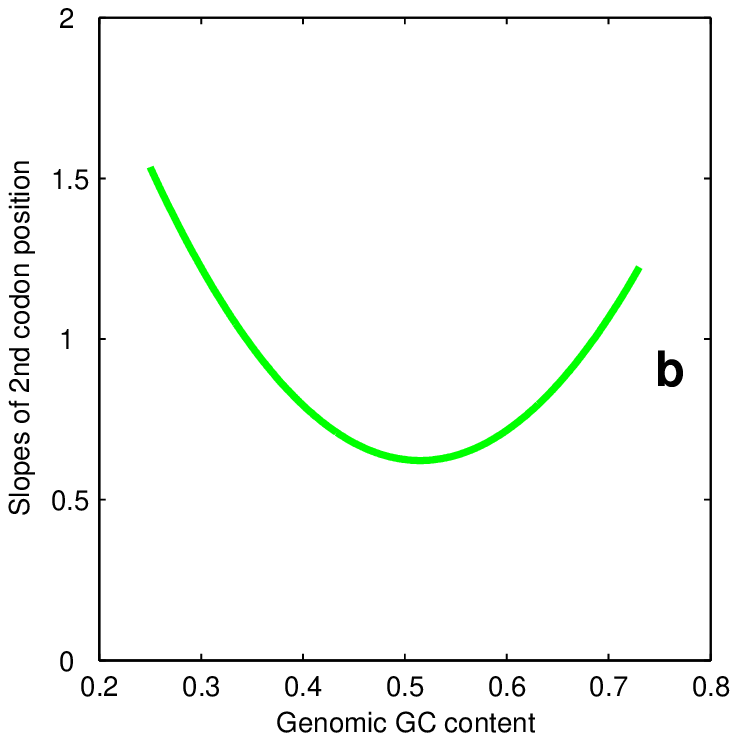}
\includegraphics[width=40mm]{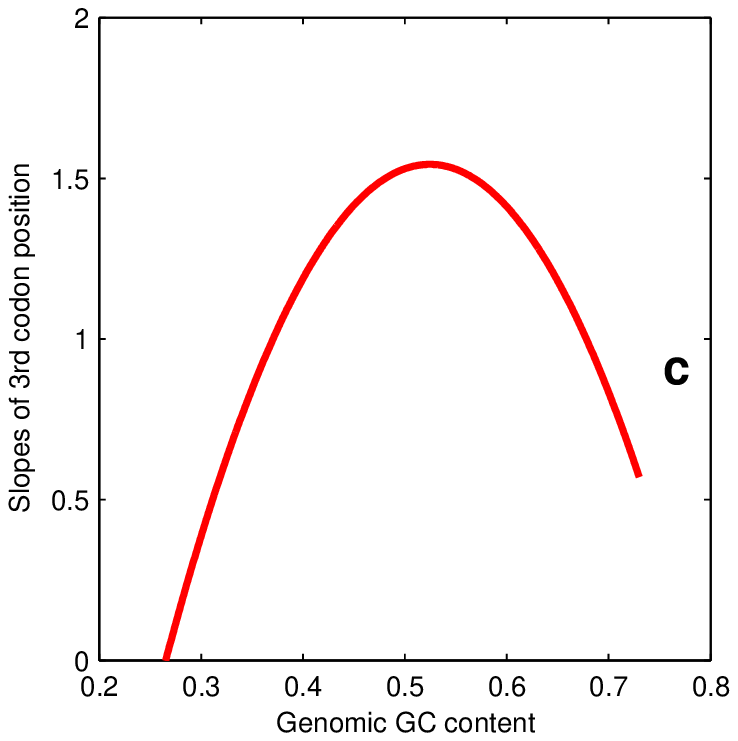}\\
\includegraphics[width=40mm]{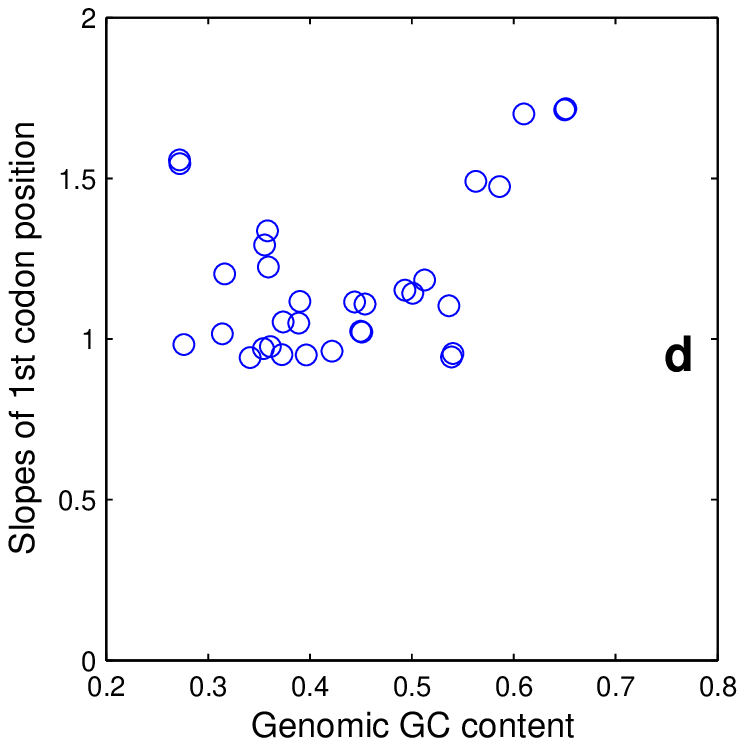}
\includegraphics[width=40mm]{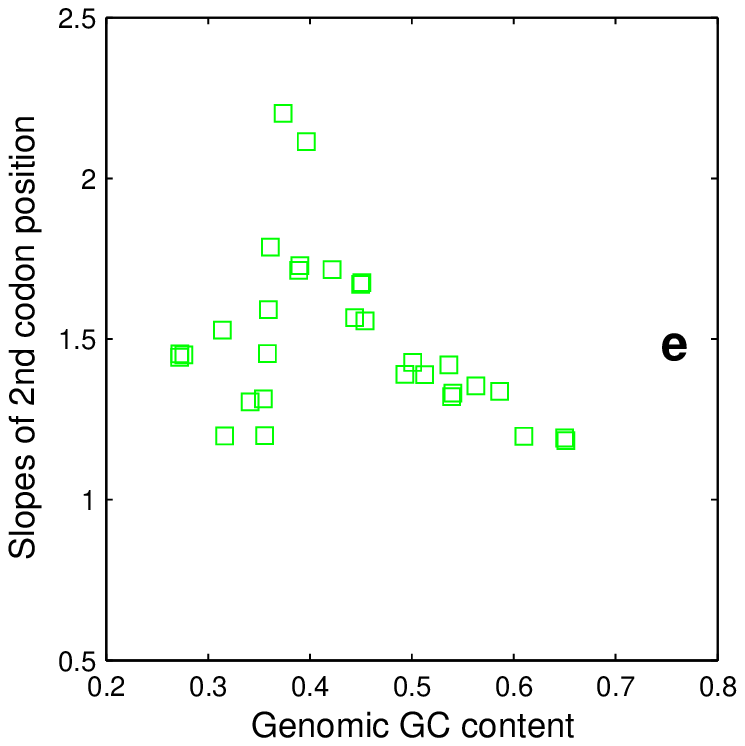}
\includegraphics[width=40mm]{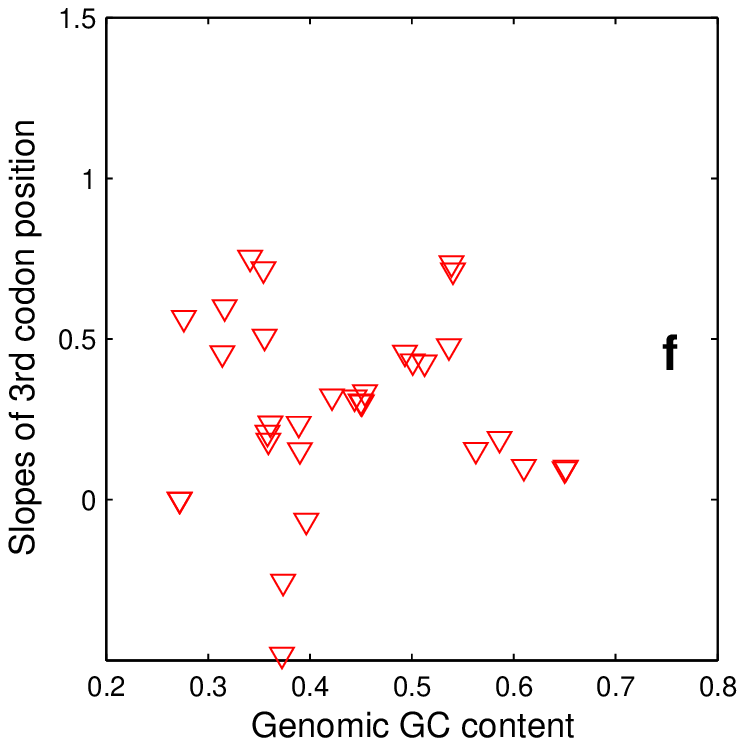}
} \label{fig} \caption{\small {\bf Fig. S10 The relationship between
the genomic GC content and slopes of genic codon plots.} The genic
codon plots for a species present the correlation between GC content
of genes in its genome and GC content at first, second and third
codon positions of the genes in its genome. {\bf a through c,} The
experimental observations for the first, second and third codon
positions respectively. {\bf d through f,} The simulations for the
first, second and third codon positions respectively, which agree
with the experimental observations in general. The bending
directions for first and third codon positions are the same with
experimental observations.}
\end{figure}

\clearpage

\begin{figure}
\centering{
\includegraphics[width=120mm]{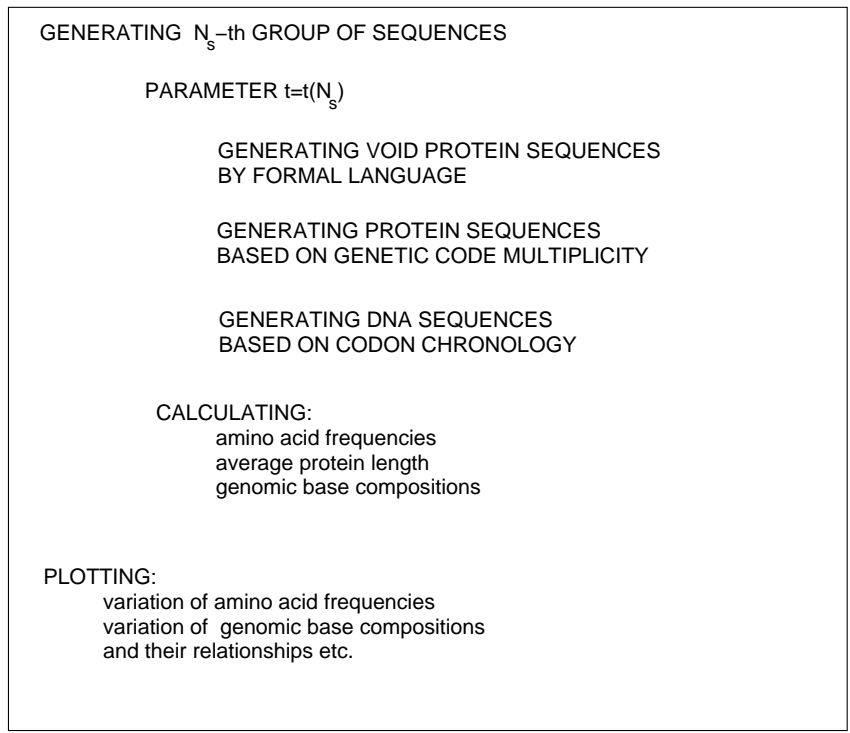}
} \label{fig} \caption{\small {\bf Fig. S11 The outline of the
program in the model.} }
\end{figure}

\clearpage

\begin{figure}
\centering{
\includegraphics[width=120mm]{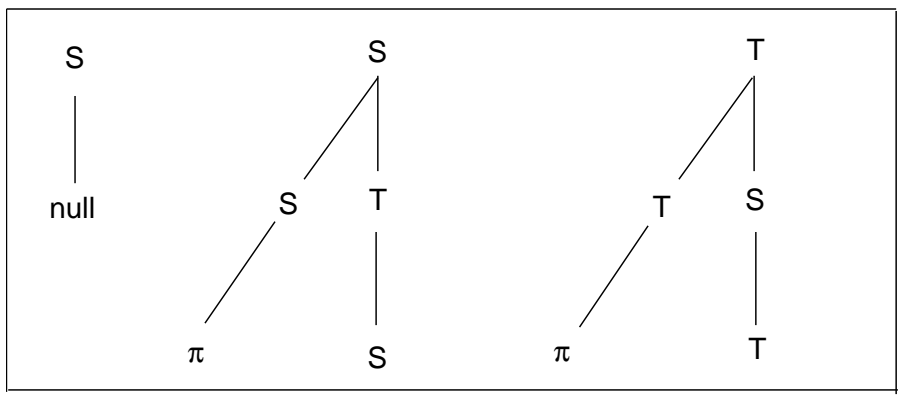}
} \label{fig} \caption{\small {\bf Fig. S12 The tree adjoining
grammar rules in the model.} There are one initial tree and two
auxiliary trees in the grammar. An example of substitutions by the
grammar rules is given in Fig. S14. $\pi$ in the trees are leaves,
which will be replaced by amino acids according to Tab. 3.}
\end{figure}

\clearpage

\begin{figure}
\centering{
\includegraphics[width=160mm]{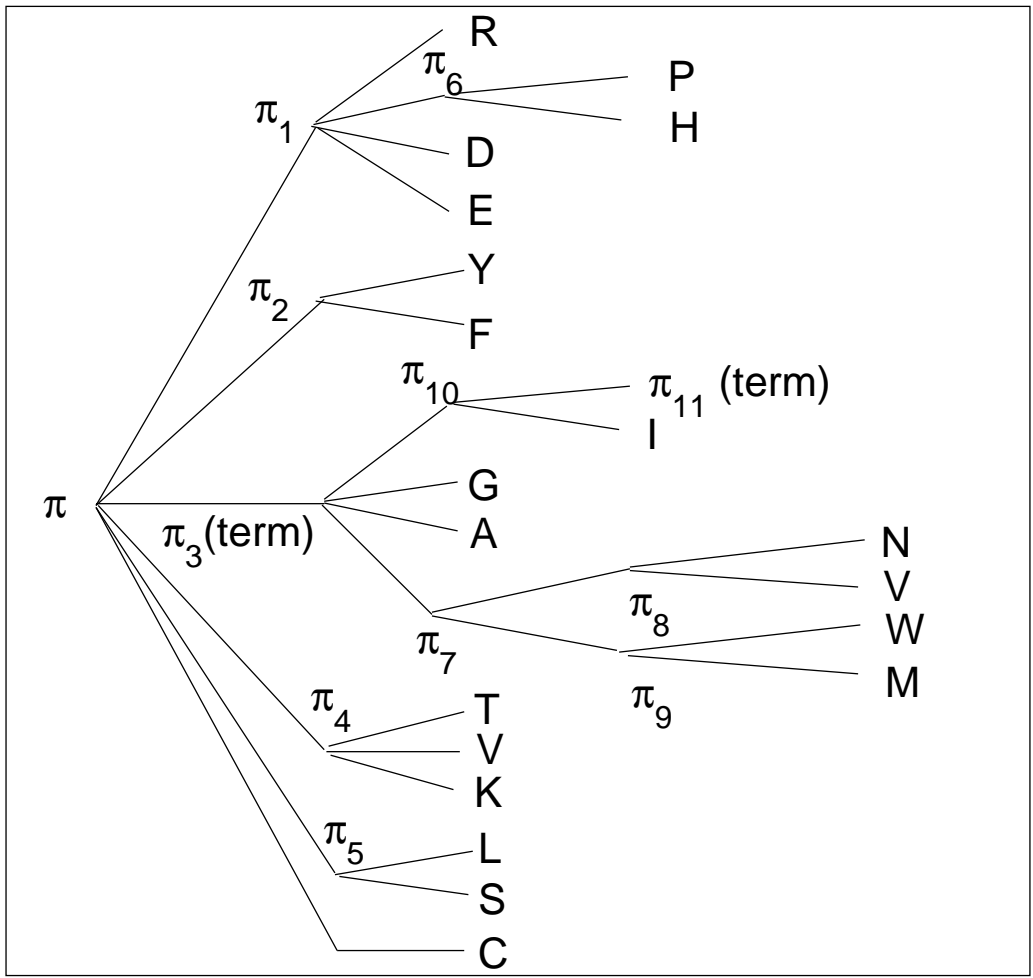}
} \label{fig} \caption{\small {\bf Fig. S13 The genetic code
multiplicity tree.} This tree is based on the genetic code
multiplicity. The probabilities for substitutions by this tree is in
Tab. 3}
\end{figure}

\clearpage

\begin{figure}
\centering{
\includegraphics[width=160mm]{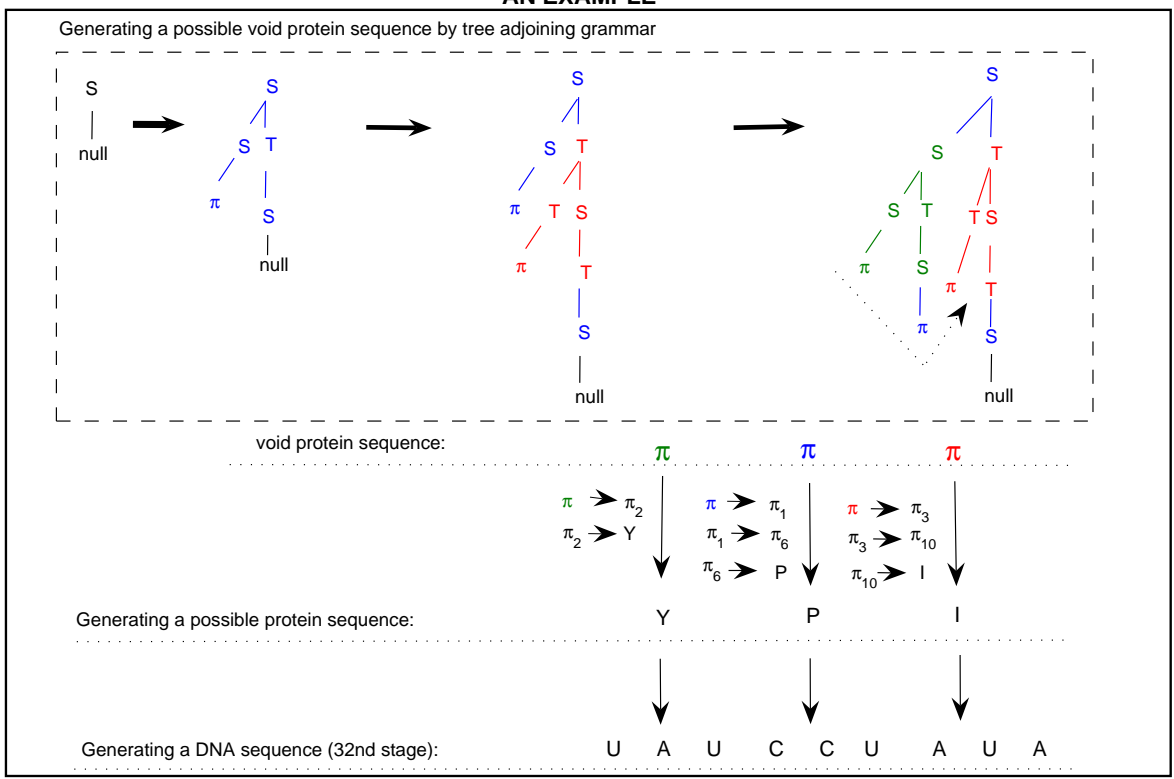}
} \label{fig} \caption{\small {\bf Fig. S14 An example of generating
protein sequences and DNA sequences by the model.} }
\end{figure}

\clearpage

\begin{center}
{\Large SUPPLEMENTARY TABLES 1-6}
\end{center}

\clearpage
\begin{sidewaystable}
  \centering
  \caption{Linear incorrelation between variation trends of amino acid frequencies and gain-loss of amino acids in modern time} \label{}

  \begin{tabular*}{22.75cm}{@{\extracolsep{-7pt}}|l|@{\hspace{4mm}}c|c|c|c|c|c|c|c|c|c|c|c|c|c|c|c|c|c|c|c|}
    \hline
&   G &  A  &  D  &  V  &  P  &  S  &  E  &  L  &  T  &  R  &    Q &
I  &   N &   H &   K &   C &   F &   Y &   M &   W \\ \hline Evol.
trend ($\times 10^{-6}$) & -354& -765& -62.9& -245 &-292 & 128&
77.0& -61.0 & -42.9& -475  & 9.95 &590 & 490 &-57.0& 734& -14.7 &
214 & 187 &-1.27& -59.5 \\ \hline Gain-loss ($\times 10^{-4}$) & -63
& -239 &-39 & 98&-139 &167 & -137 & -17 & 91 & 38 &20 &89 &73 &73
&-65 & 67 & 42&-5 & 88&2
\\    \hline
  \end{tabular*}
\end{sidewaystable}

\clearpage
\begin{sidewaystable}
  \centering \small
  \caption{The amino acid frequencies as initial values in the simulations of variation of amino acid frequencies}\label{1}

\begin{tabular*}{21.38cm}{@{\extracolsep{-3.1pt}}|l|@{\hspace{3mm}}r|r|r|r|r|r|r|r|r|r|r|r|r|r|r|r|r|r|r|r|}
  \hline
  & $f_G$ & $f_A$ & $f_D$ & $f_V$ &
  $f_P$ & $f_S$ & $f_E$ & $f_L$ &
  $f_T$ & $f_R$ & $f_Q$ & $f_I$ &
  $f_N$ & $f_H$ & $f_K$ & $f_C$ &
  $f_F$ & $f_Y$ & $f_M$ & $f_W$ \\\hline

  $iAAF_{\mbox{\tiny PEP}}$ ($\times 10^{-4}$) &692 &  831 &  524 &  674 &  489 &  703 &  647 &  991 &  537 &  559 &
394 &  577 &  400 &  226 &  555 &  141 &  404 &  299 &  238 & 119
\\\hline

  $iAAF_{\mbox{\tiny NCBI}}$ ($\times 10^{-4}$) &721 &  906 &  536 &  698 &  427 &  607 &  624 &  1025 &  527 &  537 &
366 &  672 &  413 &  204 &  560 &  97 &  414 &  312 &  238 & 116
\\\hline

  $iAAF_{\mbox{\tiny medium}}$ ($\times 10^{-4}$) & 639 & 804 & 516 & 619 & 410 & 622 & 597 & 1051 & 464 & 520 & 365 &
811 & 421 & 195 & 721 & 84 & 464 & 348 & 237 & 114 \\\hline

\end{tabular*}
\end{sidewaystable}

\clearpage

\begin{sidewaystable}
  \centering \small
  \caption{The genetic code multiplicity and substitution probability of amino acids}\label{1}

\begin{tabular}{|l|l|l|l|}
  \hline
  $\pi  \rightarrow \pi_1 $ & $\pi_1  \rightarrow R \ [p_R/p_1]$  &   &   \\
  \ $[p_1=p_D+p_R$  & $\pi_1  \rightarrow E \ [p_E/p_1]$  &   &   \\
  \ \ \ \ \ \ \ $+p_E+p_6,$  & $\pi_1  \rightarrow D \ [p_D/p_1]$  &   &   \\ \cline{3-4}
  \ \ $p_6=p_P+p_H]$  & $\pi_1  \rightarrow \pi_6 \ [p_6/p_1]$  & $\pi_6  \rightarrow H \ [p_H/p_6]$  &   \\
    &   & $\pi_6  \rightarrow P \ [p_P/p_6]$  &   \\ \cline{2-4}
  $\pi  \rightarrow \pi_2 $  & $\pi_2  \rightarrow Y \ [p_Y/p_2]$  &   &   \\
  \ $[p_2=p_F+p_Y]$  & $\pi_2  \rightarrow F \ [p_F/p_2]$  &   &   \\ \cline{2-4}
     & $\pi_3  \rightarrow \pi_{10} \ [p_I/p_3]$  & $\pi_{10}  \rightarrow \pi_{11} \ [1-\kappa]$  & $\pi_{11}  \rightarrow \pi_3 $  [1]\\ \cline{4-4}
  $\pi  \rightarrow \pi_3 $  &   & $\pi_{10}  \rightarrow I \ [\kappa]$  &   \\ \cline{3-4}
  \ $[p_3=p_I+p_A$  & $\pi_3  \rightarrow A \ [p_A/p_3]$  &   &   \\
  \ \ \ \ \ \ \ $+p_G+p_7,$  & $\pi_3  \rightarrow G \ [p_G/p_3]$  &   &   \\ \cline{3-4}
  \ \ $p_7=p_N+p_Q$  & $\pi_3  \rightarrow \pi_7 \ [p_7/p_3]$  & $\pi_7  \rightarrow \pi_8 \ [p_8/p_7,$  & $\pi_8  \rightarrow N \ [p_N/p_8]$  \\
  \ \ \ \ \ \ \ $+p_M+p_W]$  &   & \ $p_8=p_N+p_Q]$  & $\pi_8  \rightarrow Q \ [p_Q/p_8]$  \\ \cline{4-4}
    &   & $\pi_7  \rightarrow \pi_9 \ [p_9/p_3,$  & $\pi_9  \rightarrow W \ [p_W/p_9]$  \\
    &   & \ $p_9=p_M+p_W]$  & $\pi_9  \rightarrow M \ [p_M/p_9]$  \\ \cline{2-4}
 $\pi  \rightarrow \pi_4 $   & $\pi_4  \rightarrow T \ [p_T/p_4]$  &   &   \\
 \ $[p_4=p_V+p_T$   & $\pi_4  \rightarrow K \ [p_K/p_4]$  &   &   \\
  \ \ \ \ \ \ \ \ $+p_K]$  & $\pi_4  \rightarrow V \ [p_V/p_4]$  &   &   \\ \cline{2-4}
 $\pi  \rightarrow \pi_5 $   & $\pi_5  \rightarrow L \ [p_L/p_5]$  &   &   \\
 \ $[p_5=p_S+p_L]$   & $\pi_5  \rightarrow S \ [p_S/p_5]$  &   &   \\ \cline{2-4}
 $\pi  \rightarrow C \ [p_C]$   &   &   &   \\  \hline
\end{tabular}
\end{sidewaystable}

\clearpage
\begin{sidewaystable}[htbp]
  \centering
  \small
  \caption{The codon chronology based on the amino acid chronology and complementarity.}\label{}
   \begin{tabular*}{22.4cm}{@{\extracolsep{-10pt}}|r|@{\hspace{4.5mm}}cccccccccccccccccccccccc|}
    \hline
     & G &  A  &  D  &  V  &  P  &  S  &  E  &  L  &  T  &  R  &  Q  &  I  &   N &   H &   K &   C &   F &   Y &   M &   W &   (stop) &   (S) &   (L) &   (R) \\ \hline
     1 & GGC & GCC &     &     &     &     &     &     &     &     &     &     &     &     &     &     &     &     &     &
    &&&&\\ \hline
     2 &     &     & GAC & GUC &     &     &     &     &     &     &     &     &     &     &     &     &     &     &     &
 &&&&\\ \hline
     3 & GGG &     &     &     & CCC &     &     &     &     &     &     &     &     &     &     &     &     &     &     &
   &&&&\\ \hline
     4 & GGA &     &     &     &     & UCC &     &     &     &     &     &     &     &     &     &     &     &     &     &
  &&&&\\ \hline
     5 &     &     &     &     &     &     & GAG & CUC &     &     &     &     &     &     &     &     &     &     &     &
&&&&\\ \hline
     6 & GGU &     &     &     &     &     &     &     & ACC &     &     &     &     &     &     &     &     &     &     &
&&&&\\ \hline
     7 &     & GCG &     &     &     &     &     &     &     & CGC &     &     &     &     &     &     &     &     &     &
  &&&&\\ \hline
     8 &     &     &     &     & CCG &     &     &     &     & CGG &     &     &     &     &     &     &     &     &     &
&&&&\\ \hline
     9 &     &     &     &     &     & UCG &     &     &     & CGA &     &     &     &     &     &     &     &     &     &
&&&&\\ \hline
    10 &     &     &     &     &     &     &     &     & ACG & CGU &     &     &     &     &     &     &     &     &     &
&&&&\\ \hline
    11 &     &     &     &     &     &     &     & CUG &     &     & CAG &     &     &     &     &     &     &     &     &
&&&&\\ \hline
    12 &     &     & GAU &     &     &     &     &     &     &     &     & AUC &     &     &     &     &     &     &     &
&&&&\\ \hline
    13 &     &     &     & GUU &     &     &     &     &     &     &     &     & AAC &     &     &     &     &     &     &
&&&&\\ \hline
    14 &     &     &     &     &     &     &     &     &     &     &     & AUU & AAU &     &     &     &     &     &     &
&&&&\\ \hline
    15 &     &     &     & GUG &     &     &     &     &     &     &     &     &     & CAC &     &     &     &     &     &
&&&&\\ \hline
    16 &     &     &     &     &     &     &     & CUU &     &     &     &     &     &     & AAG &     &     &     &     &
&&&&\\ \hline
    17 &     & GCA &     &     &     &     &     &     &     &     &     &     &     &     &     & UGC &     &     &     &
&&&&\\ \hline
    18 &     &     &     &     &     &     &     &     & ACA &     &     &     &     &     &     & UGU &     &     &     &
&&&&\\ \hline
    19 &     &     &     &     &     &     & GAA &     &     &     &     &     &     &     &     &     & UUC &     &     &
&&&&\\ \hline
    20 &     &     &     &     &     &     &     &     &     &     &     &     &     &     & AAA &     & UUU &     &     &
&&&&\\ \hline
    21 &     &     &     & GUA &     &     &     &     &     &     &     &     &     &     &     &     &     & UAC &     &
&&&&\\ \hline
    22 &     &     &     &     &     &     &     &     &     &     &     & AUA &     &     &     &     &     & UAU &     &
&&&&\\ \hline
    23 &     &     &     &     &     &     &     &     &     &     &     &     &     & CAU &     &     &     &     & AUG &
&&&&\\ \hline
    24 &     &     &     &     & CCA &     &     &     &     &     &     &     &     &     &     &     &     &     &     & UGG
&&&&\\ \hline
    25 &     &     &     &     &     &     &     & CUA &     &     &     &     &     &     &     &     &     &     &     &
& UAG &&&\\ \hline
    26 &     &     &     &     &     & UCA &     &     &     &     &     &     &     &     &     &     &     &     &     &
& UGA &&&\\ \hline
    27 &     & GCU &     &     &     &     &     &     &     &     &     &     &     &     &     &     &     &     &     &
&& AGC &&\\ \hline
    28 &     &     &     &     &     &     &     &     & ACU &     &     &     &     &     &     &     &     &     &     &
&& AGU &&\\ \hline
    29 &     &     &     &     &     &     &     &     &     &     & CAA &     &     &     &     &     &     &     &     &
&&& UUG &\\ \hline
    30 &     &     &     &     &     &     &     &     &     &     &     &     &     &     &     &     &     &     &     &
& UAA && UUA &\\ \hline
    31 &     &     &     &     & CCU &     &     &     &     &     &     &     &     &     &     &     &     &     &     &
&&&& AGG \\ \hline
    32 &     &     &     &     &     & UCU &     &     &     &     &     &     &     &     &     &     &     &     &     &
&&&& AGA \\ \hline

  \end{tabular*}

\end{sidewaystable}

\clearpage

\begin{sidewaystable}
  \centering \small
  \caption{The modified codon chronology.}\label{}
  \begin{tabular*}{21.9cm}{@{\extracolsep{-5.2pt}}|r|@{\hspace{2mm}}c|c|c|c|c|c|c|c|c|c|c|c|c|c|c|c|c|c|c|c|}
    \hline
     & G &  A  &  D  &  V  &  P  &  S$^*$  &  E  &  L  &  T  &  R$^*$  &  Q  &  I  &   N &   H &   K &   C &   F &   Y &   M &   W \\ \hline
     1 & GGC & GCC & GAC & GUC & CCC & AGC & GAG & CUC & ACC & AGG & CAG & AUC & AAC & CAC & AAG & UGC & UUC & UAC & AUG & UGG
    \\
     2 & GGG & GCG & GAC & GUC & CCC & AGC & GAG & CUC & ACC & AGG & CAG & AUC & AAC & CAC & AAG & UGC & UUC & UAC & AUG & UGG
 \\
     3 & GGG & GCG & GAU & GUU & CCC & AGC & GAG & CUC & ACC & AGG & CAG & AUC & AAC & CAC & AAG & UGC & UUC & UAC & AUG & UGG
   \\
     4 & GGA & GCG & GAU & GUU & CCG & AGU & GAG & CUC & ACC & AGG & CAG & AUC & AAC & CAC & AAG & UGC & UUC & UAC & AUG & UGG
  \\
     5 & GGU & GCG & GAU & GUU & CCG & AGU & GAG & CUC & ACC & AGG & CAG & AUC & AAC & CAC & AAG & UGC & UUC & UAC & AUG & UGG
 \\
     6 & GGU & GCG & GAU & GUU & CCG & AGU & GAA & CUG & ACC & AGG & CAG & AUC & AAC & CAC & AAG & UGC & UUC & UAC & AUG & UGG
\\
     7 & GGU & GCG & GAU & GUU & CCG & AGU & GAA & CUG & ACG & AGG & CAG & AUC & AAC & CAC & AAG & UGC & UUC & UAC & AUG & UGG
  \\
     8 & GGU & GCA & GAU & GUU & CCG & AGU & GAA & CUG & ACG & AGA & CAG & AUC & AAC & CAC & AAG & UGC & UUC & UAC & AUG & UGG
  \\
     9 & GGU & GCA & GAU & GUU & CCA & AGU & GAA & CUG & ACG & CGC & CAG & AUC & AAC & CAC & AAG & UGC & UUC & UAC & AUG & UGG
 \\
    10 & GGU & GCA & GAU & GUU & CCA & UCC & GAA & CUG & ACG & CGG & CAG & AUC & AAC & CAC & AAG & UGC & UUC & UAC & AUG & UGG
\\
    11 & GGU & GCA & GAU & GUU & CCA & UCC & GAA & CUG & ACA & CGA & CAG & AUC & AAC & CAC & AAG & UGC & UUC & UAC & AUG & UGG
\\
    12 & GGU & GCA & GAU & GUU & CCA & UCC & GAA & CUU & ACA & CGA & CAA & AUC & AAC & CAC & AAG & UGC & UUC & UAC & AUG & UGG
\\
    13 & GGU & GCA & GAU & GUU & CCA & UCC & GAA & CUU & ACA & CGA & CAA & AUU & AAC & CAC & AAG & UGC & UUC & UAC & AUG & UGG
 \\
    14 & GGU & GCA & GAU & GUG & CCA & UCC & GAA & CUU & ACA & CGA & CAA & AUU & AAU & CAC & AAG & UGC & UUC & UAC & AUG & UGG
 \\
    15 & GGU & GCA & GAU & GUG & CCA & UCC & GAA & CUU & ACA & CGA & CAA & AUA & AAU & CAC & AAG & UGC & UUC & UAC & AUG & UGG
 \\
    16 & GGU & GCA & GAU & GUA & CCA & UCC & GAA & CUU & ACA & CGA & CAA & AUA & AAU & CAU & AAG & UGC & UUC & UAC & AUG & UGG
 \\
    17 & GGU & GCA & GAU & GUA & CCA & UCC & GAA & CUA & ACA & CGA & CAA & AUA & AAU & CAU & AAA & UGC & UUC & UAC & AUG & UGG
 \\
    18 & GGU & GCU & GAU & GUA & CCA & UCC & GAA & CUA & ACA & CGA & CAA & AUA & AAU & CAU & AAA & UGU & UUC & UAC & AUG & UGG
 \\
    19 & GGU & GCU & GAU & GUA & CCA & UCC & GAA & CUA & ACU & CGA & CAA & AUA & AAU & CAU & AAA & UGU & UUC & UAC & AUG & UGG
  \\
    20 & GGU & GCU & GAU & GUA & CCA & UCC & GAA & CUA & ACU & CGA & CAA & AUA & AAU & CAU & AAA & UGU & UUU & UAC & AUG & UGG
  \\
    21 & GGU & GCU & GAU & GUA & CCA & UCC & GAA & CUA & ACU & CGA & CAA & AUA & AAU & CAU & AAA & UGU & UUU & UAC & AUG & UGG
   \\
    22 & GGU & GCU & GAU & GUA & CCA & UCC & GAA & CUA & ACU & CGA & CAA & AUA & AAU & CAU & AAA & UGU & UUU & UAU & AUG & UGG
   \\
    23 & GGU & GCU & GAU & GUA & CCA & UCC & GAA & CUA & ACU & CGA & CAA & AUA & AAU & CAU & AAA & UGU & UUU & UAU & AUG & UGG
   \\
    24 & GGU & GCU & GAU & GUA & CCA & UCC & GAA & CUA & ACU & CGA & CAA & AUA & AAU & CAU & AAA & UGU & UUU & UAU & AUG & UGG
   \\
    25 & GGU & GCU & GAU & GUA & CCU & UCC & GAA & CUA & ACU & CGA & CAA & AUA & AAU & CAU & AAA & UGU & UUU & UAU & AUG & UGG
 \\
    26 & GGU & GCU & GAU & GUA & CCU & UCC & GAA & UUG & ACU & CGA & CAA & AUA & AAU & CAU & AAA & UGU & UUU & UAU & AUG & UGG
   \\
    27 & GGU & GCU & GAU & GUA & CCU & UCG & GAA & UUG & ACU & CGA & CAA & AUA & AAU & CAU & AAA & UGU & UUU & UAU & AUG & UGG
  \\
    28 & GGU & GCU & GAU & GUA & CCU & UCA & GAA & UUG & ACU & CGA & CAA & AUA & AAU & CAU & AAA & UGU & UUU & UAU & AUG & UGG
\\
    29 & GGU & GCU & GAU & GUA & CCU & UCU & GAA & UUG & ACU & CGA & CAA & AUA & AAU & CAU & AAA & UGU & UUU & UAU & AUG & UGG
 \\
    30 & GGU & GCU & GAU & GUA & CCU & UCU & GAA & UUA & ACU & CGA & CAA & AUA & AAU & CAU & AAA & UGU & UUU & UAU & AUG & UGG
  \\
    31 & GGU & GCU & GAU & GUA & CCU & UCU & GAA & UUA & ACU & CGA & CAA & AUA & AAU & CAU & AAA & UGU & UUU & UAU & AUG & UGG
 \\
    32 & GGU & GCU & GAU & GUA & CCU & UCU & GAA & UUA & ACU & CGU & CAA & AUA & AAU & CAU & AAA & UGU & UUU & UAU & AUG & UGG
\\
    \hline
  \end{tabular*}

\end{sidewaystable}

\clearpage
\begin{sidewaystable}
  \centering \small
  \caption{The number of bases at codon positions based on the modified codon chronology}\label{}
  \begin{tabular}{|r|ccc|ccc|ccc|cccc|cccc|}
    \hline
    & \multicolumn{3}{c|}{G} & \multicolumn{3}{c|}{C} & \multicolumn{3}{c|}{U} &
    \multicolumn{4}{c|}{GC}  &    \multicolumn{4}{c|}{CU} \\ \cline{2-18}
   & 1st & 2nd & 3rd & 1st& 2nd& 3rd  & 1st& 2nd& 3rd & 1st& 2nd& 3rd& total & 1st& 2nd& 3rd& total  \\\hline
  1& 5& 5& 6& 4 & 3 & 14 & 4 & 5 & 0 & 9 & 8 & 20 & 37 & 8 & 8 & 14 & 30 \\
  2& 5& 5& 8& 4 & 3 & 12 & 4 & 5 & 0 & 9 & 8 & 20 & 37 & 8 & 8 & 12 & 28 \\
  3& 5& 5& 8& 4 & 3 & 10 & 4 & 5 & 2 & 9 & 8 & 18 & 35 & 8 & 8 & 12 & 28 \\
  4& 5& 5& 8& 4 & 3 & 9 & 4 & 5 & 2 & 9 & 8 & 17 & 34 & 8 & 8 & 11 & 27 \\
  5& 5& 5& 8& 4 & 3 & 8 & 4 & 5 & 4 & 9 & 8 & 16 & 33 & 8 & 8 & 12 & 28 \\
  6& 5& 5& 8& 4 & 3 & 7 & 4 & 5 & 4 & 9 & 8 & 15 & 32 & 8 & 8 & 11 & 27 \\
  7& 5& 5& 9& 4 & 3 & 7 & 4 & 5 & 4 & 9 & 8 & 16 & 33 & 8 & 8 & 11 & 27 \\
  8& 5& 5& 7& 4 & 3 & 7 & 4 & 5 & 4 & 9 & 8 & 14 & 31 & 8 & 8 & 11 & 27 \\
  9& 5& 5& 6& 5 & 3 & 8 & 4 & 5 & 4 & 10 & 8 & 14 & 32 & 8 & 8 & 12 & 28 \\
  10& 5& 4& 7& 5 & 4 & 7 & 5 & 5 & 3 & 10 & 8 & 14 & 32 & 10 & 9 & 10 & 29 \\
  11& 5& 4& 5& 5 & 4 & 7 & 5 & 5 & 3 & 10 & 8 & 12 & 30 & 10 & 9 & 10 & 29 \\
  12& 5& 4& 3& 5 & 4 & 7 & 5 & 5 & 4 & 10 & 8 & 10 & 28 & 10 & 9 & 11 & 30 \\
  13& 5& 4& 3& 5 & 4 & 6 & 5 & 5 & 5 & 10 & 8 & 9 & 27 & 10 & 9 & 11 & 30 \\
  14& 5& 4& 4& 5 & 4 & 5 & 5 & 5 & 5 & 10 & 8 & 9 & 27 & 10 & 9 & 10 & 29 \\
  15& 5& 4& 4& 5 & 4 & 5 & 5 & 5 & 4 & 10 & 8 & 9 & 27 & 10 & 9 & 9 & 28 \\
  16& 5& 4& 3& 5 & 4 & 4 & 5 & 5 & 5 & 10 & 8 & 7 & 25 & 10 & 9 & 9 & 28 \\
  17& 5& 4& 2& 5 & 4 & 4 & 5 & 5 & 4 & 10 & 8 & 6 & 24 & 10 & 9 & 8 & 27 \\
  18& 5& 4& 2& 5 & 4 & 3 & 5 & 5 & 6 & 10 & 8 & 5 & 23 & 10 & 9 & 9 & 28 \\
  19& 5& 4& 2& 5 & 4 & 3 & 5 & 5 & 7 & 10 & 8 & 5 & 23 & 10 & 9 & 10 & 29 \\
  20& 5& 4& 2& 5 & 4 & 2 & 5 & 5 & 8 & 10 & 8 & 4 & 22 & 10 & 9 & 10 & 29 \\
  21& 5& 4& 2& 5 & 4 & 2 & 5 & 5 & 8 & 10 & 8 & 4 & 22 & 10 & 9 & 10 & 29 \\
  22& 5& 4& 2& 5 & 4 & 1 & 5 & 5 & 9 & 10 & 8 & 3 & 21 & 10 & 9 & 10 & 29 \\
  23& 5& 4& 2& 5 & 4 & 1 & 5 & 5 & 9 & 10 & 8 & 3 & 21 & 10 & 9 & 10 & 29 \\
  24& 5& 4& 2& 5 & 4 & 1 & 5 & 5 & 9 & 10 & 8 & 3 & 21 & 10 & 9 & 10 & 29 \\
  25& 5& 4& 2& 5 & 4 & 1 & 5 & 5 & 10 & 10 & 8 & 3 & 21 & 10 & 9 & 11 & 30 \\
  26& 5& 4& 3& 5 & 4 & 1 & 6 & 5 & 10 & 10 & 8 & 4 & 22 & 11 & 9 & 11 & 31 \\
  27& 5& 4& 4& 5 & 4 & 0 & 6 & 5 & 10 & 10 & 8 & 4 & 22 & 11 & 9 & 10 & 30 \\
  28& 5& 4& 3& 5 & 4 & 0 & 6 & 5 & 10 & 10 & 8 & 3 & 21 & 11 & 9 & 10 & 30 \\
  29& 5& 4& 3& 4 & 4 & 0 & 6 & 5 & 11 & 9 & 8 & 3 & 20 & 10 & 9 & 11 & 30 \\
  30& 5& 4& 2& 4 & 4 & 0 & 6 & 5 & 11 & 9 & 8 & 2 & 19 & 10 & 9 & 11 & 30 \\
  31& 5& 4& 2& 4 & 4 & 0 & 6 & 5 & 11 & 9 & 8 & 2 & 19 & 10 & 9 & 11 & 30 \\
  32& 5& 4& 2& 4 & 4 & 0 & 6 & 5 & 12 & 9 & 8 & 2 & 19 & 10 & 9 & 12 & 31 \\
 \hline
  \end{tabular}

\end{sidewaystable}


\begin{thebibliography}{99}

\bibitem{Cell review} Knight R. D., and Landweber L. F. The early
evolution of the genetic code. Cell 101, 569-572 (2000).

\bibitem{genetic code} Szathm\'{a}ry, E. Why are there four letters
in the genetic alphabet? Nature Rev. Genetics 4, 995-1001, (2003).

\bibitem{genetic code evol} Crick, F. H. C. The Origin of the
Genetic Code. J. Mol. Biol. 38, 376-379 (1968).

\bibitem{genetic code evol2} Osawa, S. et al., Recent evidence for
evolution of the genetic code. Microbiol. Rev. {\bf 56} 229-264
(1992).

\bibitem{genetic code evol3} Trifonov, E. N. et al., Distinct
stages of protein evolution as suggested by protein sequence
analysis. {\it J. Mol. Evol.} {\bf 53}, 394-401 (2001).

\bibitem{PEP} Carter, P., Liu, J. and Rost, B. PEP: Predictions
for Entire Proteomes. Nucleic Acids Research 31, 410-413 (2003).

\bibitem{GC data} Haft D. H., Selengut J. D., Brinkac L. M., Zafar N.,
and White O. Genome Properties: a system for the investigation of
prokaryotic genetic content for microbiology, genome annotation and
comparative genomics. Bioinformatics 21, 293-306 (2005).

\bibitem{Forsdyke} Forsdyke, D. R., {\it Evolutionary Bioinformatics}
(Springer, New York, 2006).

\bibitem{PRL multiplicity} Hornos, J. E. M., and Hornos, Y. M. M.,
Algebraic Model for the Evolution of the Genetic Code. {\it Phy.
Rev. Lett.} {\bf 71}, 4401-4404 (1993).

\bibitem{gain_loss} Jordan, I. K. et al. A univeral trend of
amino acid gain and loss in protein evolution. Nature 433, 633-638,
(2005).

\bibitem{aaf constant} Rost, B. Did evolution leap to create the
protein universe? Curr. Opin. Stru. Biol. 12, 409-416 (2002).

\bibitem{aaf constant2} Liu, J.-F. and Rost, B. Comparing function
and structure between entire proteomes. Protein Sci. 10, 1970-1979
(2001).

\bibitem{AA Chronology} Trifonov, E. N. The triplet code from first
principle. J. Biomol. Struct. Dyn. 22, 1-11 (2004)

\bibitem{Savitzky-Golay} Savitzky, A. and Golay, M. J. E. Smoothing
and differentiation of data. {\it Anal. Chem.} {\bf 36}, 1627-1639
(1964).

\bibitem{3domains} Woese, C. R., Kandler, O. and Wheelis, M. L.
Towards a natural system of organisms: Proposal for the domains
Archaea, Bacteria, and Eucarya. Proc. Natl. Acad. Sci. USA 87,
4576-4579 (1990).

\bibitem{GC GC123} Muto A., and Osawa S. The guanine and cytosine content of
genomic DNA and bacterial evolution. Proc. Natl. Acad. Sci. USA 84,
166-169 (1987).

\bibitem{GC GC123 physica a} Gorban, A. et al., Codon usage trajectories and
7-cluster structure of 143 complete bacterial genomic sequences.
{\it Physica A} {\bf 353}, 365-387 (2005).

\bibitem{genetic code chronology} Trifonov, E. N., Consensus temporal order
of amino acids and evolution of the triplet code. Gene {\bf 261},
139-151 (2000). 53, 394-401 (2001).

\bibitem{aaf GC} Sueoka N. Correlation between base composition of
deoxyribonucleic acid and amino acid composition of protein. Proc.
Natl. Acad. Sci. USA 47, 1141-1149 (1961).

\bibitem{aaf GC2} Gu X., Hewett-Emmett D., and Li W.-H.
Directional mutational pressure affects the amino acid composition
and hydrophobicity of proteins in bacteria. Genetica 103, 383-391
(1998).

\bibitem{gain_loss debate} Hurst, L. D., Feil E. J., Rocha E. P. C.
Causes of trends in amino-acid gain and loss. Nature 442, E11-E12
(2006).

\bibitem{TAG} Joshi, A. K. and Schabes, Y., in Handbook of Formal Lan-
guages, eds G. Rozenberg and A. Salomma, pp.69-214 (Springer,
Heidelberg, 1997).

\bibitem{genetic code chronology2} Trifonov, E. N., Kirzhner, A., Kirzhner,
V. M., and Berezovsky, I. N. Distinct stages of protein evolution as
suggested by protein sequence analysis. J. Mol. Evol. 53, 394-401
(2001).

\bibitem{complementarity} Eigen, M., and Schuster, P., The
hypercycle. A principle of natural self-organization. Part C: The
realistic hypercycle. Naturwissenschaften {\bf 65} 341-369 (1978).

*E-mail: dirson@mail.xjtu.edu.cn





\clearpage
\end{thebibliography}
\end{document}